\documentclass[usenatbib]{mn2e}

\usepackage{threeparttable}
\usepackage{times}
\usepackage{epsfig}
\usepackage{textcomp}
\usepackage{amssymb}
\usepackage{subfigure}
\usepackage{multirow}
\usepackage{amsmath}
\usepackage{pdflscape}
\usepackage{graphicx}
\usepackage{appendix}

\begin{document}

\title[X-ray properties of BzK-selected galaxies]{X-ray properties of BzK-selected galaxies in the deepest X-ray fields}
\author[C. Rangel et al.]{C.~Rangel$^1$, K.~Nandra$^2$, E.~S.~Laird$^1$, P.~Orange$^1$\\
$^1$Astrophysics Group, Imperial College London, Blackett Laboratory, Prince Consort Road, London SW7 2AZ\\
$^2$Max-Planck-Institut f\"{u}r extraterrestrische Physik, Giessenbachstrasse 1, D-85748 Garching bei M\"{u}nchen, Germany}

\maketitle

\begin{abstract}
We investigate the X-ray properties of BzK-selected galaxies at $z\sim2$ using deep X-ray data in the Chandra Deep Field South (CDFS) and Chandra Deep Field North (CDFN). A subset of these BzK galaxies have been proposed as Compton thick Active Galactic Nuclei (AGN) candidates based on a high ratio of infrared (IR) and ultraviolet (UV) star formation rates (SFR). With the benefit of deep 24$\mu$m observations our sample of these IR excess galaxies is larger than previous studies, and combined with the deepest X-ray data yield new insights into the population. We identify 701 and 534 star forming BzKs (sBZKs) in the range $z=1.2-3.0$ in the CDFS and CDFN respectively. Of these we directly detect in X-rays 49 sBzKs in CDFS and 32 sBzKs in CDFN. Stacking the undetected sources also reveals a significant X-ray signal. Investigating the X-ray detection rate and stacked flux versus the IR excess parameter (i.e. SFR$_{total}$/SFR$_{UV, corr}$), we find no strong evidence for an increased X-ray detection rate, or a harder X-ray spectrum in IR Excess sBzKs. This is particularly the case when one accounts for the strong correlation between the IR excess parameter and the bolometric IR luminosity ($L_{IR}$), e.g. when controlling for $L_{IR}$, the  IR Non-Excess sBzKs show a detection rate at least as high. While both direct detections and stacking suggest that the AGN fraction in sBzK galaxies is high, there is no clear evidence for widespread Compton thick activity in either the sBzK population generally, or the IR Excess sBzK subsample. The very hard X-ray signal obtained for the latter in earlier work was most likely contaminated by a few hard X-ray sources now directly detected in deeper X-ray data. The X-ray detection fraction of passive BzK galaxies in our sample is if anything higher than that of sBZKs,  so there is no evidence for coeval black hole growth and star formation from X-ray analysis of the BzK populations.  Because increased AGN activity in the IR excess population is not indicated by our X-ray analysis, it appears that the bulk of the IR Excess sBzK population are luminous star-forming galaxies whose SFRs are either overestimated at 24$\mu$m, underestimated in the UV, or both. This conclusion reinforces recent results from {\it Herschel} which show similar effects. 

\end{abstract}

\section{Introduction}
\label{sec:intro}

Extensive deep multi-wavelength observations across the sky have vastly improved our knowledge of star-formation and AGN activity across a range of redshifts. It has become abundantly clear that the Universe at $z\sim 1-3$, was much more active than it is currently, at $z\sim0$. For example, the star-formation rate density steadily rises from $z=0$ to $z=1$, flattens out until at least $z\sim$3 and then falling off again at the highest redshifts \citep{lilly96,madau96,hopkins05}. AGN activity shows a similar form, indicating that globally, at least, the growth of supermassive black holes roughly tracks the buildup of the stellar population in galaxies (e.g. \citealt{boyle93}; \citealt{silverman08}; \citealt{aird10}). Observing galaxies at z$\sim$2 should give us tremendous insight into this evolution, and the interplay between AGN and star-formation in their host galaxies during this more active phase, but spectroscopic identifications, and hence our understanding of galaxies with $1.4\le z<2.5$ is complicated by the ``redshift desert". At these redshifts there are very few spectral features at observed-frame optical wavelengths that can be used to determine the redshift. In response to this problem numerous selection techniques have been developed that use broad band photometry instead of spectroscopy to select $z\sim$2 candidate galaxies. For example, the Lyman-break and Balmer break features have been used to select galaxies with $1<z<3$ \citep{erb03,adelberger04}.

The BzK selection technique of \citet{daddi04} combines optical ($B$ and $z$ band) and near-IR ($K$ band) photometry to target the Balmer and 4000~\AA~break features of star-forming and passively evolving galaxies at $z\sim$2 \citep{reddy05,shapley05,wuyts08}. The selection technique is expected to be largely independent of reddening \citep{daddi04} and the near-IR photometry allows extremely red passively evolving galaxies to be selected, as well as bluer, star-forming galaxies \citep{thompson99,franx03,mccarthy04}. BzK selection shows good agreement with other $z\sim$2 galaxy selection techniques, although it is liable to miss the bluest/youngest star-forming galaxies (K$_S>21$ Vega mag; \citealt{reddy05}). Many sBzKs show strong emission in the mid-IR at 24$\mu$m, the majority of which is attributable to dusty star formation \citet{daddi07a}. By comparing UV and IR derived star-formation rates (SFR's) \citet{daddi07b} identified sBzKs with excess IR emission (IR Excess sBzK) as candidates for containing AGN. \citet{daddi07b} to attempted to isolate sBzKs which might contain AGN using this mid-IR excess. AGN are expected to exhibit strong continuum emission at mid-IR wavelengths due to re-radiated emission from hot dust (e.g. \citealt{pier92}; \citealt{efstathiou95}), a feature that has been exploited previously to identify obscured AGN \citep{lacy04,stern05}.

\citet{daddi07b} stacked the X-ray emission of IR Excess sBzKs in the GOODS-S field, yielding a hard X-ray signal ($\Gamma\sim0.8^{+0.4}_{-0.3}$) consistent with a significant population of Compton thick (N$_H>1.5\times10^{24}$ cm$^{-2}$) AGN. The stacked signal of X-ray undetected sBzKs that did not exhibit excessive IR emission (IR Non-Excess sBzKs) appeared softer ($\Gamma\sim1.7$) and consistent with pure star-formation. These results have significant implications for co-eval star-formation and black hole growth and the composition of the X-ray background. The presence of ``buried" Compton thick AGN in star-forming galaxies (like sBzKs) has been predicted by models of AGN-galaxy-co-evolution (e.g. \citealt{hopkins05}). The majority of the X-ray background \citep{giacconi62} has been resolved using the emission of discrete X-ray sources (as predicted by \citealt{setti89}), but there remains an unexplained hard (E $>$ 6 keV) X-ray deficit \citep{worsley05}. Models of the X-ray background predict the hard X-ray background deficit could represent a large population of as yet undiscovered Compton thick AGN at $z\sim1-2$ \citep{comastri95,gilli07}. Extrapolating the stacked X-ray emission of the X-ray undetected IR-Excess sBzKs over all the sky would explain a substantial portion ($\sim10\%-25\%$) of the hard X-ray background deficit \citep{daddi07b}. \citet{alexander11} carried out further analysis of sBzKs using new, deeper 4Ms CDFS X-ray data and found that stacking X-ray spectra of heavily obscured ($\Gamma\le1$) sBzKs produced a hard, reflection dominated spectrum consistent with Compton thick AGN emission. The stacked spectrum features a large EW Iron line at rest-frame E $\sim$ 6.4 keV similar to that observed in Compton thick AGN in the local Universe \citep{matt96,iwasawa98} and distant Universe \citep{georgantopoulos09,comastri11,feruglio11,gilli11}. \citet{alexander11} stacked the spectra of sBzK galaxies containing heavily obscured AGN and found their spectra was best fit by a pure reflection model with a strong iron K$\alpha$ line at rest-frame 6.4 keV. This is consistent with Compton thick absorption (N$_H\sim10^{24}$cm$^-2$) in a small fraction of sBzK galaxies. Stacking X-ray undetected sBzKs in deeper 4Ms CDFS data did not yield conclusive evidence of a high fraction of heavily obscured AGN in the more general sBzK population, however, somewhat in disagreement with the earlier work of  \citet{daddi07b}. These observations imply that the space density of Compton thick AGN at $z\sim2$ is comparable to that of unobscured AGN at $z\sim2$.

The IR excess analysis of \citet{daddi07b} is reliant upon the accuracy of IR and UV SFR estimates. Mid-IR observations have been found to be reliable proxies for bolometric IR luminosities \citep{dale05} and it has been common practice to fit 24$\mu$m observations (rest-frame 8$\mu$m at $z\sim2$) with empirically derived star-forming galaxy templates (e.g. \citealt{chabaz01}; \citealt{dalou02}) to estimate the SFR. With the advent of \textit{Herschel} our understanding of the Far-IR spectra of star-forming galaxies has improved markedly, especially at $z>1$. Using Far-IR \textit{Herschel} data \citet{elbaz11} calculated $L_{IR}$ for numerous galaxies with MIPS coverage and compared them to $L_{IR}$ estimates made using 24$\mu$m selected star-forming templates of \citet{chabaz01}. Based on this analysis it is concluded that the star-forming templates of \citet{chabaz01} are overestimating the bolometric IR luminosity for galaxies with $L_{IR}>10^{12}$ $L_{\odot}$. Additionally \citet{nordon12} find that the IR Excess emission at rest-frame 8$\mu$m can be explained entirely by PAH features, removing the necessity for mid-IR continuum emission from AGN assumed by previous IR selection techniques. The UV estimates of SFR are known to suffer in dusty (ie heavily star-forming) galaxies because reddening corrections become less accurate as the dust opacity increases \citep{goldader02}. Thus, selection of AGN via comparison of IR and UV-derived SFRs may not be as robust as previously thought. Indeed, recent work have suggested that some IR excess selection techniques (e.g. \citealt{fiore08}) have been shown to select equal parts heavily obscured AGN, low-luminosity AGN and purely star-forming galaxies \citep{donley08,georgakakis10} when followed up using X-ray data. 

In the light of this, in the present paper we revisit the issue of the AGN content in IR Excess BzK galaxies pioneered by \citet{daddi07b}. We examine the X-ray properties of IR luminous sBzKs, selected in a similar manner to \citet{daddi07b}, using 4Ms CDFS and 2Ms CDFN X-ray data, through a mixture of direct detections and stacking. In Section \ref{sub:optir} we discuss the optical and IR data used in this work. Section \ref{sub:sample} outlines the BzK and 24$\mu$m selection methodology while Sections \ref{sub:xray_redux} and \ref{sub:stacking} outline the X-ray data reduction and X-ray stacking technique respectively. In Section \ref{sec:results} we present the results of direct X-ray detection of sBzKs and stacking of X-ray undetected sBzKs and in Section \ref{sec:comparison} we compare our findings to those of \citet{daddi07b} and \citet{alexander11}. In Section \ref{sec:discussion} we discuss the implications of these results and in Section \ref{sec:summary} we summarise our findings. Throughout this work a standard, flat $\Lambda$CDM cosmology with $\Omega_\Lambda = 0.7$ and H$_0 = 70$~km~s$^{-1}$~Mpc$^{-1}$ is assumed.  

\section{Data}
\label{sec:data}
\subsection{Optical and IR data}
\label{sub:optir}
In the CDFS we use \textit{Hubble} ACS F435W, F606W, F775W and F850LP band (henceforth $B, v, i$ and $z$ respectively), ISAAC $K$ band, IRAC 3.6$\mu$m and MIPS 24$\mu$m band data from the GOODS-MUSIC-v2 catalogue (\citealt{santini09}). All objects have either a spectroscopic (64\%) or a photometric (36\%) redshift. This dataset presents identical $B, v, i, z, K$ and 3.6$\mu$m photometry to the GOODS-MUSIC catalogue (\citealt{grazian06}) that was used by \citet{daddi07b} but incorporates deeper MIPS 24$\mu$m band data from the FIDEL survey (PI: Mark Dickinson). The 24$\mu$m source fluxes in GOODS-MUSIC-v2 were identified and matched to the multi-wavelength data using the ConvPhot software \citep{desantis07}, as opposed to matching the MIPS sources to multi-wavelength data using a more basic fixed matching radius method (as was the case for \citealt{daddi05} and subsequently Daddi et al 2007a and 2007b). The additional depth and use of the ConvPhot software allows the identification of additional 24$\mu$m sources and therefore increases our sample size compared to \citet{daddi07b}. The GOODS-MUSIC-v2 catalogue has more spectroscopic redshifts than GOODS-MUSIC, improving the accuracy of our luminosity calculations. In the CDFN we use \textit{Hubble} ACS $B, V, i$ and $z$ band, MOIRCS $K_S$, IRAC 3.6$\mu$m and MIPS 24$\mu$m band photometry collected for the MOIRCS Deep Survey (MODS; \citealt{kajisawa11}). MODS is the only publicly available catalogue in CDFN to date to have matched K band to 24$\mu$m sources. Again, sources have a mixture of spectroscopic (70\%) and photometric (30\%) redshifts.

\begin{figure}
\includegraphics[width=80mm]{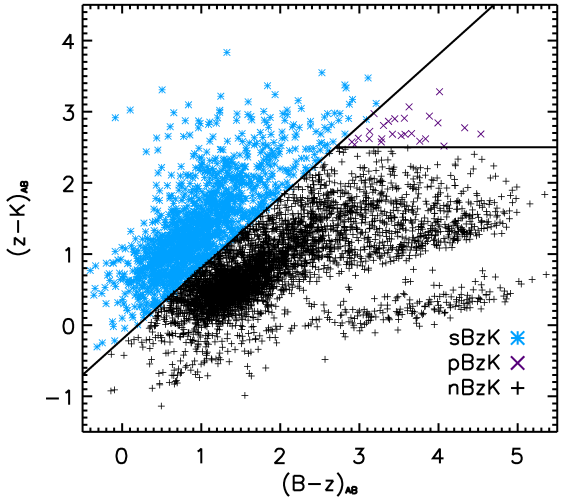}

\caption{BzK diagram of galaxies in the GOODS-MUSIC-v2 and MODS catalogues. Solid Black lines represent the boundaries between different BzK galaxy type criteria.}
\label{fig:bzk}
\end{figure}

To ensure the reliability of the BzK selection and IR excess analysis we require all the optically galaxies in our sample to satisfy the following criteria:
\begin{enumerate}
  \item A secure K-band detection ($K_S<22$ Vega).
  \item less than 0.5 arcseconds separation from an IRAC 3.6um selected source to verify that the source is genuine (large PSFs of 24$\mu$m sources leads to source blending).
  \item Error on $(B-z)_{AB}<0.4$ to ensure reddening corrections are reliable.
  \item Either a spectroscopic or photometric redshift.
\end{enumerate}
In total 5,685 galaxies across CDFS and CDFN satisfy these criteria (plotted on BzK diagram; Figure \ref{fig:bzk}). 

\begin{table}
\label{tab:bzkclass}
\begin{center}
\caption{BzK galaxy samples and IR Excess and Non-Excess sBzK subsamples for the CDFS and CDFN.}
\begin{tabular}{@{}lccc}
\hline
Galaxy type & Total CDFS & Total CDFN & Total Combined\\
\hline
pBzK & 21 & 5 & 26\\
nBzK & 2168 & 2006 & 4174\\
sBzK & 851 & 594 & 1445\\
IRX ($1.2\le z\le3.0$) & 158 & 198 & 356\\
IRNX ($1.2\le z\le3.0$) & 543 & 336 & 879\\
\hline
\end{tabular}
\end{center}
\end{table}

\begin{figure}
\includegraphics[width=80mm]{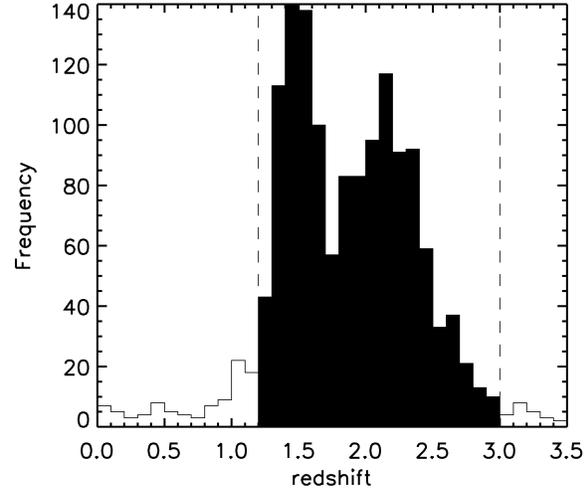}

\caption{Redshift distribution of all 1485 sBzK galaxies. Filled portion of the histogram represents sBzK galaxies subjected to IR Excess analysis ($1.2\le z\le3.0$), the remaining galaxies are omitted as they possess either $z<1.2$ or $z>3$.}
\label{fig:z_dist}
\end{figure}

\begin{figure}
\includegraphics[width=80mm]{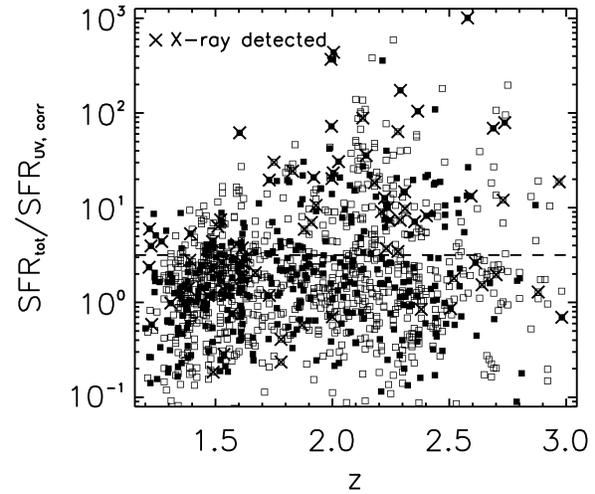}

\caption{Plot of SFR$_{tot}$/SFR$_{UVcorr}$ versus redshift for sBzK galaxies used for IR excess analysis. Galaxies above the horizontal dashed line satisfy the IR excess galaxy criterion, galaxies below are IR non-excess. Filled squares are galaxies with spectroscopic redshifts while open squares are galaxies with photometric redshift estimates only. Black crosses represent galaxies that have been directly detected in the 4Ms CDFS and 2Ms CDFN data.}
\label{fig:xs_vs_z}
\end{figure}

\begin{figure}
\includegraphics[width=80mm]{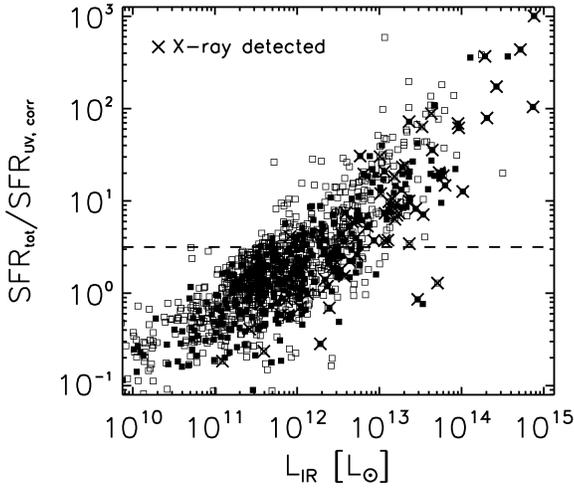}

\caption{Plot of SFR$_{tot}$/SFR$_{UVcorr}$ versus $L_{IR}$ for sBzK galaxies used for IR excess analysis. Filled squares are galaxies with spectroscopic redshifts while open squares are galaxies with photometric redshift estimates only. The horizontal dashed line denotes the IR excess criterion; galaxies above the line are IR excess, below the line are IR non-excess. Black crosses represent galaxies that have been directly detected in the 4Ms CDFS and 2Ms CDFN data.}
\label{fig:xs_vs_lir}
\end{figure}

\subsection{IR Excess sBzK selection}
\label{sub:sample}
BzK galaxies with excess IR emission compared to that expected for star formation, may have a significant contribution to the IR flux originating from accretion and are therefore candidate obscured AGN. To define IR Excess sBzK galaxies we follow the methods of \citet{daddi07b}, using a combination of the BzK selection technique \citep{daddi04} and SFR's calculated using UV and IR observations. The star-forming BzK (sBzK) selection (Equation \ref{eq:sbzk}) of \citet{daddi04} identifies star-forming galaxies at $z>1.4$.
\begin{equation}
\label{eq:sbzk}
(z-K)_{AB} - (B-z)_{AB} > -0.2
\end{equation}
The $z-K$ colour acts a proxy for stellar mass, probing the optical SED about the Balmer and 4000~\AA~breaks for galaxies with $z\sim2$ \citep{reddy05,shapley05,wuyts08}, while the $B-z$ colour corrects for reddening effects (see Equation \ref{eq:ebv}). Additionally the sBzK selection was empirically calibrated to make it independent of reddening effects (\citealt{daddi04}). This is an important feature because emission from regions of intense star-formation is expected to be heavily reddened. Only galaxies that satisfy the sBzK criterion are subjected to 24$\mu$m excess analysis. The BzK plot (see Figure \ref{fig:bzk}) also identifies passively evolving galaxies at z$\sim$2 (pBzK) and normal galaxies at $z<1.4$ (nBzK), but these are not included in the subsequent IR Excess analysis. While the sBzK criterion primarily selects $z\sim2$ galaxies there remains a fraction of low redshift interlopers (see Figure \ref{fig:z_dist}). Additionally, a large spread in redshift distribution would also adversely affect the X-ray stacking analysis. Therefore the redshift range for sBzKs subjected to IR Excess analysis is narrowed to $1.2\le z<3.0$. The results of BzK selection and the subsequent IR Excess analysis are outlined in Table 1.

For the purpose of this analysis we calculate SFR$_{IR}$ following the relationship \citet{kennicutt98}:
\begin{equation}
\label{eq:sfrir}
SFR_{IR}(M_{\odot} yr^{-1}) = 1.73 \times10^{-10}L_{IR}(L_{\odot})
\end{equation}
where $L_{IR}$ is the bolometric IR luminosity of the galaxy between 8 and 1000 $\mu$m ($L_{8-1000\mu}$$_m$). The observed 24$\mu$m flux is k-corrected to rest-frame 8$\mu$m flux using the IR star-forming galaxy templates of \citet{chabaz01}. The \citet{chabaz01} templates are empirically derived from observations of local star-forming galaxies. \citet{magnelli09} found that the \citet{chabaz01} templates exhibit a stronger correlation to the emission of observed galaxies at $z\sim1$ than the star-forming galaxy templates of \citet{dalou02} and \citet{lagache03}. The bolometric $L_{IR}$ is inferred from monochromatic 8$\mu$m luminosity ($L_{8um}$) using the relationships for \citet{chabaz01} templates defined in \citet{daddi07b}.
\begin{equation}
log(\frac{L_{IR}}{L_{\odot}}) = 1.50log(\frac{L_{8\mu m}}{L_{\odot}}) - 4.31
\end{equation}
if $log(\frac{L_{8\mu m}}{L_{\odot}}) > 9.75$ and
\begin{equation}
log(\frac{L_{IR}}{L_{\odot}}) = 0.93log(\frac{L_{8\mu m}}{L_{\odot}}) + 1.23
\end{equation}
if $log(\frac{L_{8\mu m}}{L_{\odot}}) < 9.75$.

SFR$_{UV}$ is estimated using Equation \ref{eq:sfruv}, a relationship derived by \citet{daddi04}:
\begin{equation}
\label{eq:sfruv}
SFR(M_{\odot}yr^{-1})=\frac{L_{1500}(erg s^{-1} Hz^{-1})}{8.85 \times 10^{27}}
\end{equation}
where $L_{1500}$ is the 1500 \AA~monochromatic luminosity. The UV spectrum is approximated to a power law derived from observed $B, v, i$ and $z$ bands (rest-frame UV at $z\sim2$) and $L_{1500}$ is then calculated. The dust reddening factor $E(B-V)$ is estimated using the following criterion.
\begin{equation}
\label{eq:ebv}
E(B-V)=0.25(B-z+0.1)_{AB}
\end{equation}
The UV emission is corrected for reddening, following the empirically derived reddening correction method of \citet{calzetti00}, and we use this reddening corrected UV emission to again calculate the SFR (SFR$_{UV,corr}$). To be defined as IR Excess sBzKs (hereafter IRXs; Compton thick AGN candidates) a galaxy must satisfy Equation \ref{eq:xs_crit}.
\begin{equation}
\label{eq:xs_crit}
log(\frac{SFR_{UV}+SFR_{IR}}{SFR_{UVcorr}}) > 0.5
\end{equation}
The sBzKs that do not satisfy this criterion are classified as IR Non-Excess sBzKs (hereafter IRNXs).

If an apparent IRX (see Equation \ref{eq:xs_crit}) possesses a 24$\mu$m flux that is less than 3 times the background noise it is not considered a secure-detection and removed from the sample, as the measurement of an IR excess is not necessarily robust. Clearly it is neither necessary nor desirable to apply this cut to IRNXs. Between CDFS and CDFN there are 356 IRXs and 879 IRNXs. A K-S test demonstrates that the IRXs and IRNXs have significantly different redshift distributions, with null hypothesis p$<$0.001. The IRXs have higher redshifts on average (median $z=2.06$) compared to the IRNXs (median $z=1.78$), see Figure \ref{fig:xs_vs_z}. The high density of spectroscopic observations at $z=1.6$ is due to a cluster in the CDFS sample that has been extensively observed \citep{castellano07}. There is also a strong positive correlation between $L_{IR}$ and the ratio of $SFR_{tot}/SFR_{UV,corr}$ (the IR excess parameter) in Figure \ref{fig:xs_vs_lir}. As a consequence, the overwhelming majority of IR luminous sBzKs are classified as IRXs (85.3\%; 304/356 with $L_{IR}>10^{12}$$L_{\odot}$) as opposed to IRNXs (19.1\%; 168/879 with $L_{IR}>10^{12}$$L_{\odot}$).

\subsection{X-ray data reduction}
\label{sub:xray_redux}
The 4Ms CDFS~\textit{Chandra} data was obtained over the course of 3 separate Chandra observing cycles: 1 Megasecond (Ms) between May 27th 2000 and December 23rd 2000, another 1Ms between September 20th 2007 and November 4th 2007 and the final 2Ms between May 18th 2010 and July 22nd 2010. The 2Ms CDFN data was collected between 23rd February 2000 and February 22nd 2002. All observations were taken using the Advanced CCD Imaging Spectrometer imaging array (ACIS-I; \citealt{garmire03}).

We now present a brief description of the data reduction method which incorporates the use of CIAO data analysis software version 4.2 tools and custom software. For a more detailed description of the method see \citet{laird09} and \citet{nandra}. The raw data for each observation (henceforth, obsID) is corrected for known aspect offsets and hot pixels are removed. Then charge transfer inefficiency (CTI) and time dependant gain corrections are applied. Potential flare events are identified and removed in each obsID using custom software (for a more details see \citealt{nandra07}). We manually removed flares from 5 obsID's (441, 581, 1431, 2405 and 2406) in CDFS and 2 obsID's (966 and 3389) in CDFN.

An additional astrometric correction to each obsID is calculated by comparing X-ray source positions to those of a reference source catalogue with superior astrometry. Running the CIAO wavdetect tool on each obsID, bright X-ray sources are selected using a threshold probability $10^{-6}$. These should be common to the majority of individual obsIDs and hence ideal for marrying up the astrometry. In CDFS we take reference astrometry from a catalogue of H-band selected galaxies (sources with the H-band magnitudes between 18 and 23 and no photometric flags) from the GOODS-MUSYC survey (\citealt{gawiser06}). We removed the brightest sources to eliminate extended objects with potentially incorrect positions, and faint magnitudes were omitted to reduce the likelihood of chance associations. The reference astrometry for 2Ms CDFN is from an r-band selected catalogue of \citet{capak04}. We did not apply a magnitude cut to the r-band catalogue because it made no practical difference to the astrometry correction.

Following the astrometry correction, merged images and exposure maps are created in four energy bands; 0.5-2 keV (soft band), 0.5-7 keV (full band), 2-7 keV (hard band) and 4-7 keV (ultra-hard band). Photons with energies $<0.5$ keV and $>7$ keV are ignored because the response of the ACIS-I instrument drops off markedly at these energies \citep{weisskopf02}. Using the methods of \citet{laird09} , point source catalogues containing flux estimates and PSF's are generated for the 4Ms CDFS and 2Ms CDFN merged images. The full, hard and ultra-hard band fluxes are corrected to 0.5-10 keV, 2-10 keV and 5-10 keV respectively using a power-law model with $\Gamma=1.4$ and galactic N$_H=8.95\times10^{19}$ cm$^{-2}$ and N$_H=1.51\times10^{20}$ cm$^{-2}$ in CDFS and CDFN respectively.

Our point source lists for 4Ms CDFS and 2Ms CDFN images were generated following the method of \citet{laird09}, finding 569 and 504 X-ray point sources respectively. The threshold poisson probability for sources in this catalogue is $4\times10^{-6}$ (a more in depth description of the detection method can be found in \citealt{laird09}) to ensure a low false identification rate. The 4Ms CDFS source catalogue of \citet{xue11} contains 740 sources, of which 500 are coincident with sources from our catalogue when matched using a fixed 1.5 arcsecond circular radius (corresponding false match probability 1.0\%). The \citet{xue11} catalogue has more point sources because of the lower threshold in their analysis. The 69 sources unique to our catalogue are low significance detections according to our methodology. Likewise, the 240 sources unique to the \citet{xue11} catalogue are low significance detections according to their method, the majority of which have poisson probabilities just below our detection threshold.

\subsection{Stacking Procedure}
\label{sub:stacking}
We infer the average X-ray emission of the X-ray undetected sBzKs by stacking them. Stacking involves taking a sample of X-ray faint galaxies and summing their emission to create a long exposure image with an improved Signal-to-Noise (S/N) ratio. Data collected using~\textit{Chandra} ACIS-I can be summed to create extremely long exposures without reaching the confusion limit \citep{brandt01}, so is ideal for stacking analysis. The stacking technique employed here is based on the work of \citet{nandra02} and \citet{laird05}. Fixed diameter circular apertures are used (as opposed to variable aperture techniques) to extract the source counts, as this has proved the most effective and reliable source count extraction method for X-ray stacking \citep{laird05}. A fixed aperture of radius 1.5 arcseconds provides the best S/N ratio in both 4Ms CDFS and 2Ms CDFN. Sources at large off axis angles (OAA) have wider PSFs, therefore the source counts become more diffuse and less significant relative to the background. Thus, including sources with large OAA's can have a negative effect on the S/N ratio of the final stacked image. To overcome this problem we define a maximum OAA, known as the `inclusion radius', for sources that are to be stacked. We find an inclusion radius of 5.5 arcminutes gives the optimum S/N ratio for stacks in both CDFS and CDFN.

Prior to stacking we mask out any sources that are within an area defined by the 2 times 90\% PSF of an X-ray detected source. This prevents bright sources from contaminating the stack and skewing the low count statistics of the faint sources we are interested in. A local background is estimated for each source by extracting counts in an annulus centred on the candidate source with inner and outer radii of 10 arcseconds and 30 arcseconds respectively. Pixels from which the background counts are extracted are chosen by randomly shuffling the position 10,000 times within the annulus and averaging the counts per pixel. Any part of the annulus that overlaps with the 2 times 90\% PSF of an X-ray detected source is masked out to prevent bright source contamination. X-ray undetected candidate sources within the background annulus are not masked out because they are not confirmed X-ray sources and must still be treated as part of the background. The total background counts are scaled to match the source extraction aperture in both area and exposure.

\begin{table*}
\label{tab:cdfs_direct}
\caption{X-ray properties of sBzK galaxies detected in the 4Ms CDFS. Column (1): X-ray source name, galaxies suffixed with \dag~have been detected in the 1Ms CDFS also; column (2): BzK subsample; columns (3) and (4): right ascension (RA) and declination (Dec.) of source; column (5): redshift, redshift's are spectroscopic where available (suffixed with $^s$), otherwise photometric redshifts are used (suffixed with $^p$); column (6): soft band flux, units $10^{-16}$ erg cm$^{-2}$ s$^{-1}$; column (7): hard band flux, units $10^{-16}$ erg cm$^{-2}$ s$^{-1}$; column (8): rest-frame hard band Luminosity derived from the soft band flux, units $10^{42}$ ergs s$^{-1}$; column (9): HR, HR$=(H-S)/(H+S)$, where $H$ and $S$ are hard- and soft-band count rates respectively.}
\begin{tabular}{@{} l l c c c c c c c c}
\hline
Name & Type & RA & Dec & Redshift & $F_{0.5-2 keV}$ & $F_{2-10 keV}$ & $L_{2-10 keV}$ & HR\\
(1) & (2) & (3) & (4) & (5) & (6) & (7) & (8) & (9)\\
\hline
cdfs4Ms\_031\dag & IRX & 53.15731 & -27.87007 & 1.60$^s$ & $65.69^{+1.18}_{-1.14}$ & $119.92^{+3.43}_{-3.29}$ & $88.47^{+1.59}_{-1.53}$ & $-0.43^{+0.01}_{-0.01}$ \\

cdfs4Ms\_043\dag & IRX & 53.16146 & -27.85597 & 2.97$^p$ & $0.84^{+0.15}_{-0.14}$ & $7.31^{+1.09}_{-0.90}$ & $7.78^{+1.41}_{-1.28}$ & $0.30^{+0.10}_{-0.10}$ \\

cdfs4Ms\_077\dag & IRX & 53.14879 & -27.82116 & 2.58$^s$ & $0.49^{+0.11}_{-0.10}$ & $3.52^{+0.69}_{-0.62}$ & $2.92^{+0.65}_{-0.57}$ & $0.21^{+0.13}_{-0.12}$\\

cdfs4Ms\_078\dag & IRX & 53.1801 & -27.82061 & 1.92$^s$ & $28.11^{+0.75}_{-0.72}$ & $58.78^{+2.39}_{-2.21}$ & $66.15^{+1.77}_{-1.69}$ & $-0.37^{+0.02}_{-0.02}$ \\

cdfs4Ms\_079 & IRX & 53.05903 & -27.81944 & 2.28$^p$ & $0.25^{+0.09}_{-0.08}$ & $<1.10$ & $1.02^{+0.36}_{-0.31}$ & $-0.08^{+0.31}_{-0.22}$ \\ 

cdfs4Ms\_091\dag & IRX & 53.18579 & -27.80996 & 2.59$^s$ & $3.76^{+0.29}_{-0.25}$ & $11.12^{+1.19}_{-1.02}$ & $22.68^{+1.77}_{-1.54}$ & $-0.21^{+0.06}_{-0.06}$ \\

cdfs4Ms\_120\dag & IRX & 53.18344 & -27.77658 & 2.69$^s$ & $5.56^{+0.35}_{-0.31}$ & $8.04^{+1.10}_{-0.93}$ & $37.53^{+2.37}_{-2.11}$ & $-0.51^{+0.05}_{-0.05}$ \\ 

cdfs4Ms\_123\dag & IRX & 53.04907 & -27.7745 & 2.21$^p$ & $8.52^{+0.45}_{-0.40}$ & $25.24^{+1.74}_{-1.56}$ & $31.09^{+1.62}_{-1.48}$ & $-0.21^{+0.04}_{-0.04}$ \\

cdfs4Ms\_126\dag & IRX & 53.13114 & -27.77308 & 2.22$^s$ & $0.60^{+0.12}_{-0.11}$ & $1.92^{+0.60}_{-0.55}$ & $2.22^{+0.44}_{-0.40}$ & $-0.14^{+0.16}_{-0.16}$ \\

cdfs4Ms\_131\dag & IRX & 53.09394 & -27.76773 & 1.73$^s$ & $3.26^{+0.27}_{-0.23}$ & $35.16^{+1.85}_{-1.68}$ & $5.55^{+0.46}_{-0.39}$ & $0.40^{+0.04}_{-0.04}$ \\

cdfs4Ms\_144\dag & IRX & 53.12491 & -27.75832 & 1.22$^s$ & $52.01^{+1.03}_{-0.99}$ & $124.08^{+3.44}_{-3.20}$ & $30.26^{+0.60}_{-0.57}$ & $-0.31^{+0.01}_{-0.02}$ \\

cdfs4Ms\_161\dag & IRX & 53.04549 & -27.73748 & 1.62$^s$ & $54.79^{+1.09}_{-1.05}$ & $114.69^{+3.47}_{-3.29}$ & $75.51^{+1.50}_{-1.44}$ & $-0.37^{+0.01}_{-0.02}$ \\

cdfs4Ms\_175\dag & IRX & 53.10697 & -27.71826 & 2.29$^s$ & $13.07^{+0.55}_{-0.51}$ & $55.59^{+2.53}_{-2.35}$ & $53.40^{2.26}_{-2.09}$ & $-0.04^{+0.03}_{-0.03}$ \\

cdfs4Ms\_183\dag & IRX & 53.1337 & -27.69866 & 1.39$^s$ & $6.68^{+0.43}_{-0.039}$ & $16.35^{+1.90}_{-1.72}$ & $5.85^{+0.38}_{-0.34}$ & $-0.30^{+0.05}_{-0.06}$ \\

cdfs4Ms\_215\dag & IRX & 53.15812 & -27.88552 & 2.73$^p$ & $0.91^{+0.18}_{-0.16}$ & $<1.34$ & $6.44^{+1.26}_{-1.17}$ & $-0.49^{+0.24}_{-0.24}$ \\

cdfs4Ms\_224 & IRX & 53.17406 & -27.85968 & 1.93$^p$ & $0.38^{+0.13}_{-0.12}$ & $1.08^{+0.62}_{-1.08}$ & $0.91^{+0.31}_{-0.28}$ & $-0.06^{+0.29}_{-0.22}$ \\

cdfs4Ms\_264\dag & IRX & 53.05881 & -27.70842 & 2.03$^s$ & $1.62^{+0.28}_{-0.24}$ & $7.57^{+1.70}_{-1.53}$ & $4.51^{+0.78}_{-0.66}$ & $0.02^{+0.13}_{-0.12}$ \\

cdfs4Ms\_334 & IRX & 53.16222 & -27.71213 & 2.23$^p$ & $0.50^{+0.20}_{-0.19}$ & $6.56^{+1.55}_{-1.39}$ & $1.89^{+0.76}_{-0.73}$ & $0.42^{+0.15}_{-0.15}$ \\

cdfs4Ms\_369 & IRX & 53.06313 & -27.69953 & 2.40$^s$ & $1.76^{+0.30}_{-0.25}$ & $4.25^{+1.77}_{-1.76}$ & $8.32^{+1.41}_{-1.20}$ & $-0.25^{+0.18}_{-0.15}$ \\

cdfs4Ms\_434 & IRX & 53.19878 & -27.84391 & 1.51$^p$ & $0.28^{+0.13}_{-0.12}$ & $8.64^{+1.24}_{-1.07}$ & $0.31^{+0.14}_{-0.13}$ & $0.69^{+0.09}_{-0.09}$ \\

cdfs4Ms\_438 & IRX & 53.16685 & -27.79876 & 2.00$^s$ & $0.26^{+0.09}_{-0.08}$ & $0.94^{+0.50}_{-0.54}$ & $0.68^{+0.24}_{-0.20}$ & $-0.02^{+0.27}_{-0.21}$ \\

cdfs4Ms\_451 & IRX & 53.1312 & -27.84129 & 1.61$^s$ & $<0.08$ & $1.06^{+0.50}_{-0.47}$ & $<0.11$ & $0.48^{+0.28}_{-0.18}$ \\

cdfs4Ms\_460 & IRX & 53.09755 & -27.71552 & 2.14$^s$ & $<0.14$ & $4.40^{+1.29}_{-1.16}$ & $<0.46$ & $0.76^{+0.21}_{-0.09}$ \\

cdfs4Ms\_515 & IRX & 53.05683 & -27.79852 & 1.83$^p$ & $0.14^{+0.08}_{-0.07}$ & $<0.60$ & $0.28^{+0.16}_{-0.15}$ & $-0.19^{+0.42}_{-0.35}$ \\

cdfs4Ms\_013\dag & IRNX & 53.19605 & -27.89266 & 2.74$^p$ & $5.45^{+0.40}_{-0.35}$ & $22.19^{+2.00}_{-1.92}$ & $39.08^{+2.83}_{-2.53}$ & $-0.06^{+0.06}_{-0.05}$\\

cdfs4Ms\_042\dag & IRNX & 53.15067 & -27.85735 & 1.61$^s$ & $2.15^{+0.23}_{-0.19}$ & $6.12^{+0.97}_{-0.80}$ & $2.95^{+0.31}_{-0.26}$ & $-0.22^{+0.08}_{-0.08}$ \\

cdfs4Ms\_050\dag & IRNX & 53.07463 & -27.84865 & 1.54$^s$ & $1.58^{+0.19}_{-0.18}$ & $47.66^{+2.21}_{-2.03}$ & $1.87^{0.23}_{-0.21}$ & $0.73^{+0.03}_{-0.03}$ \\

cdfs4Ms\_083\dag & IRNX & 53.04569 & -27.81555 & 1.39$^p$ & $11.49^{+0.48}_{-0.45}$ & $22.20^{+1.54}_{-1.38}$ & $9.99^{+0.42}_{-0.39}$ & $-0.40^{+0.03}_{-0.03}$ \\

cdfs4Ms\_088\dag & IRNX & 53.14987 & -27.81405 & 1.31$^s$ & $1.22^{+0.16}_{-0.15}$ & $1.98^{+0.56}_{-0.52}$ & $0.88^{+0.12}_{-0.11}$ & $-0.44^{+0.11}_{-0.11}$ \\

cdfs4Ms\_089\dag & IRNX & 53.17929 & -27.81254 & 1.73$^s$ & $1.22^{+0.17}_{-0.16}$ & $19.82^{+1.48}_{-1.31}$ & $2.07^{+0.29}_{-0.26}$ & $0.55^{+0.06}_{-0.05}$ \\

cdfs4Ms\_133\dag & IRNX & 53.16283 & -27.76716 & 1.22$^s$ & $16.02^{+0.59}_{-0.55}$ & $52.66^{+2.32}_{-2.14}$ & $9.27^{+0.34}_{-0.32}$ & $-0.23^{+0.03}_{-0.03}$ \\

cdfs4Ms\_155\dag & IRNX & 53.02409 & -27.74643 & 1.61$^s$ & $3.78^{+0.33}_{-0.29}$ & $8.01^{+1.49}_{-1.32}$ & $5.14^{+0.45}_{-0.39}$ & $-0.35^{+0.09}_{-0.07}$ \\

cdfs4Ms\_157\dag & IRNX & 53.16272 & -27.74424 & 2.52$^p$ & $5.11^{+0.35}_{-0.31}$ & $20.13^{+1.61}_{-1.43}$ & $28.13^{+1.92}_{-1.70}$ & $-0.07^{+0.05}_{-0.05}$ \\

cdfs4Ms\_206\dag & IRNX & 53.08925 & -27.93046 & 2.61$^s$ & $5.07^{+0.42}_{-0.38}$ & $14.16^{+2.32}_{-1.77}$ & $31.25^{+2.61}_{-2.33}$ & $-0.18^{+0.07}_{-0.08}$ \\

cdfs4Ms\_208\dag & IRNX & 53.20484 & -27.91799 & 2.02$^p$ & $7.21^{+0.52}_{-0.47}$ & $13.50^{+2.56}_{-2.13}$ & $19.87^{+1.44}_{-1.30}$ & $-0.41^{+0.07}_{-0.08}$ \\

cdfs4Ms\_226\dag & IRNX & 53.0601 & -27.8529 & 1.54$^s$ & $0.45^{+0.13}_{-0.12}$ & $6.83^{+1.12}_{-0.94}$ & $0.55^{+0.16}_{-0.14}$ & $0.50^{+0.12}_{-0.09}$ \\

cdfs4Ms\_246 & IRNX & 53.18171 & -27.78293 & 1.57$^s$ & $0.31^{+0.11}_{-0.10}$ & $<0.71$ & $0.39^{+0.14}_{-0.12}$ & $-0.37^{0.33}_{-0.32}$ \\

cdfs4Ms\_258 & IRNX & 53.06154 & -27.73409 & 1.67$^p$ & $1.08^{+0.19}_{-0.18}$ & $3.31^{+1.16}_{-1.06}$ & $1.65^{+0.29}_{-0.27}$ & $-0.16^{+0.17}_{-0.15}$ \\

cdfs4Ms\_271 & IRNX & 53.14107 & -27.70119 & 1.23$^s$ & $0.55^{+0.23}_{-0.22}$ & $2.27^{+1.90}_{-1.03}$ & $0.33^{+0.13}_{-0.13}$ & $0.05^{+0.29}_{-0.22}$ \\

cdfs4Ms\_283 & IRNX & 53.08729 & -27.92957 & 2.88$^p$ & $1.93^{+0.32}_{-0.28}$ & $18.48^{+2.41}_{-2.19}$ & $16.19^{+2.72}_{-2.35}$ & $0.35^{+0.08}_{-0.08}$ \\

cdfs4Ms\_321 & IRNX & 53.02796 & -27.74872 & 2.38$^p$ & $0.73^{+0.18}_{-0.17}$ & $1.99^{+1.40}_{-0.94}$ & $3.37^{+0.85}_{-0.79}$ & $-0.12^{+0.26}_{-0.19}$ \\

cdfs4Ms\_370 & IRNX & 53.05303 & -27.69705 & 1.78$^p$ & $1.51^{+0.31}_{-0.26}$ & $4.06^{+1.92}_{-1.85}$ & $2.81^{+0.57}_{-0.49}$ & $-0.19^{+0.23}_{-0.14}$ \\

cdfs4Ms\_405 & IRNX & 53.041 & -27.83608 & 1.78$^p$ & $0.20^{+0.11}_{-0.10}$ & $1.94^{+0.84}_{-0.82}$ & $0.38^{+0.20}_{-0.19}$ & $0.28^{+0.26}_{-0.18}$ \\

cdfs4Ms\_441 & IRNX & 53.09494 & -27.75793 & 1.87$^p$ & $<0.07$ & $1.60^{+0.66}_{-0.62}$ & $<0.15$ & $0.67^{+0.26}_{-0.13}$ \\

cdfs4Ms\_459 & IRNX & 53.12284 & -27.72285 & 1.63$^p$ & $<0.14$ & $3.85^{+1.18}_{-1.04}$ & $<0.20$ & $0.70^{+0.23}_{-0.12}$ \\

cdfs4Ms\_494 & IRNX & 53.1255 & -27.88646 & 2.64$^p$ & $0.66^{+0.15}_{-0.14}$ & $<1.11$ & $4.18^{+0.95}_{-0.87}$ & $-0.46^{+0.26}_{-0.26}$ \\

cdfs4Ms\_498 & IRNX & 53.12022 & -27.79886 & 1.38$^s$ & $0.28^{+0.10}_{-0.08}$ & $<0.48$ & $0.24^{+0.08}_{-0.07}$ & $-0.50^{+0.26}_{-0.32}$ \\

cdfs4Ms\_503 & IRNX & 53.16382 & -27.7735 & 2.70$^p$ & $0.15^{+0.09}_{-0.08}$ & $<0.66$ & $1.05^{+0.59}_{-0.53}$ & $-0.17^{+0.41}_{-0.33}$ \\

cdfs4Ms\_512 & IRNX & 53.11056 & -27.8236 & 1.47$^s$ & $0.24^{+0.09}_{-0.07}$ & $<0.32$ & $0.25^{+0.09}_{-0.07}$ & $-0.70^{+0.08}_{-0.30}$ \\

cdfs4Ms\_301 & pBzK & 53.24571 & -27.86106 & 1.10$^p$ & $3.02^{+0.36}_{-0.31}$ & $9.08^{+1.97}_{-1.81}$ & $1.22^{+0.14}_{-0.13}$ & $-0.18^{+0.11}_{-0.10}$ \\

cdfs4Ms\_318 & pBzK & 53.04476 & -27.77442 & 1.61$^s$ & $0.80^{+0.15}_{-0.14}$ & $<1.29$ & $1.08^{+0.21}_{-0.19}$ & $-0.45^{+0.22}_{-0.20}$ \\

cdfs4Ms\_457 & pBzK & 53.02676 & -27.76529 & 1.33$^s$ & $<0.21$ & $2.95^{+1.17}_{-1.09}$ & $<0.16$ & $0.49^{+0.25}_{-0.19}$ \\

\hline
\end{tabular}
\end{table*}

\pagebreak

\begin{table*}
\label{tab:cdfn_direct}
\caption{X-ray properties of sBzK galaxies detected in the 2Ms CDFN. Column (1): X-ray source name; column (2): BzK subsample; columns (3) and (4): right ascension (RA) and declination (Dec.) of source; column (5): column (5): redshift, redshift's are spectroscopic where available (suffixed with $^s$), otherwise photometric redshifts are used (suffixed with $^p$); column (6): soft band flux, units $10^{-16}$ erg cm$^{-2}$ s$^{-1}$; column (7): hard band flux, units $10^{-16}$ erg cm$^{-2}$ s$^{-1}$; column (8): rest-frame hard band Luminosity derived from the soft band flux, units $10^{42}$ ergs s$^{-1}$; column (9): HR, HR$=(H-S)/(H+S)$, where $H$ and $S$ are hard- and soft-band count rates respectively.}
\begin{tabular}{@{} l l c c c c c c c}
\hline
Name & Type & RA & Dec & Redshift & $F_{0.5-2 keV}$ & $F_{2-10 keV}$ & $L_{2-10 keV}$ & HR\\
(1) & (2) & (3) & (4) & (5) & (6) & (7) & (8) & (9)\\
\hline

hdfn\_015 & IRX & 189.13587 & 62.13338 & 1.99$^s$ & $1.32^{+0.31}_{-0.28}$ & $13.99^{+2.29}_{-1.94}$ & $3.49^{+0.81}_{-0.74}$ & $0.35^{+0.12}_{-0.11}$ \\

hdfn\_050 & IRX & 189.06689 & 62.18556 & 1.91$^p$ & $17.52^{+0.85}_{-0.77}$ & $149.24^{+5.29}_{-4.95}$ & $40.57^{+1.96}_{-1.79}$ & $0.28^{+0.02}_{-0.03}$ \\

hdfn\_060 & IRX & 189.22245 & 62.19449 & 1.27$^s$ & $0.77^{+0.19}_{-0.16}$ & $<1.25$ & $0.51^{+0.12}_{-0.11}$ & $-0.47^{+0.20}_{-0.20}$ \\

hdfn\_067 & IRX & 189.23252 & 62.20028 & 2.74$^s$ & $1.30^{+0.23}_{-0.21}$ & $1.53^{+0.80}_{-0.73}$ & $9.32^{+1.64}_{-1.48}$ & $-0.52^{+0.14}_{-0.15}$ \\

hdfn\_080 & IRX & 189.14395 & 62.2115 & 1.22$^s$ & $0.76^{+0.18}_{-0.16}$ & $2.44^{+0.85}_{-0.74}$ & $0.45^{+0.11}_{0.09}$ & $-0.16^{+0.18}_{-0.17}$ \\

hdfn\_082 & IRX & 189.26096 & 62.21223 & 2.31$^p$ & $2.49^{+0.31}_{-0.28}$ & $4.26^{+1.94}_{-0.93}$ & $10.46^{+1.28}_{-1.19}$ & $-0.45^{+0.09}_{-0.11}$ \\

hdfn\_098 & IRX & 189.35108 & 62.23327 & 1.59$^p$ & $5.29^{+0.51}_{-0.43}$ & $55.17^{+3.49}_{-3.12}$ & $6.95^{+0.66}_{-0.56}$ & $0.37^{+0.04}_{-0.05}$ \\

hdfn\_107 & IRX & 189.14843 & 62.24015 & 2.01$^s$ & $1.78^{+0.26}_{-0.24}$ & $20.20^{+2.02}_{-1.70}$ & $4.81^{+0.70}_{-0.64}$ & $0.39^{+0.07}_{-0.07}$ \\

hdfn\_119 & IRX & 189.21598 & 62.25136 & 2.18$^p$ & $7.10^{+0.55}_{-0.48}$ & $22.20^{+2.21}_{-1.86}$ & $24.82^{+1.92}_{-1.67}$ & $-0.21^{+0.05}_{-0.06}$ \\

hdfn\_120 & IRX & 189.25198 & 62.25248 & 2.29$^p$ & $3.31^{+0.38}_{-0.36}$ & $13.61^{+1.78}_{-1.65}$ & $13.51^{+1.56}_{-1.45}$ & $-0.07^{+0.08}_{-0.09}$ \\

hdfn\_127 & IRX & 189.34672 & 62.2607 & 2.24$^s$ & $4.42^{+0.47}_{-0.40}$ & $15.77^{+2.19}_{-1.82}$ & $16.90^{+1.81}_{-1.51}$ & $-0.14^{+0.09}_{-0.07}$ \\

hdfn\_137 & IRX & 189.27041 & 62.26714 & 1.88$^p$ & $27.65^{+1.04}_{-0.97}$ & $80.18^{+3.98}_{-3.63}$ & $60.95^{+2.30}_{-2.14}$ & $-0.25^{+0.03}_{-0.03}$ \\

hdfn\_144 & IRX & 189.31159 & 62.27141 & 1.52$^s$ & $8.93^{+0.62}_{-0.55}$ & $24.56^{+2.46}_{-2.10}$ & $10.24^{+0.72}_{-0.63}$ & $-0.27^{+0.06}_{-0.05}$ \\

hdfn\_158 & IRX & 189.17957 & 62.28656 & 2.28$^p$ & $11.17^{+0.73}_{-0.65}$ & $30.20^{+2.80}_{-2.39}$ & $44.94^{+2.94}_{-2.60}$ & $-0.28^{+0.05}_{-0.05}$ \\

hdfn\_168 & IRX & 189.42735 & 62.30341 & 2.31$^s$ & $83.05^{+1.87}_{-1.80}$ & $158.46^{+6.07}_{-5.69}$ & $347.76^{+7.85}_{-7.52}$ & $-0.43^{+0.02}_{-0.02}$ \\

hdfn\_173 & IRX & 189.3245 & 62.3155 & 2.24$^s$ & $37.72^{+1.23}_{-1.16}$ & $74.93^{+4.14}_{-3.78}$ & $143.06^{+4.68}_{-4.40}$ & $-0.41^{+0.03}_{-0.02}$ \\

hdfn\_322 & IRX & 189.36066 & 62.34105 & 2.37$^s$ & $2.29^{+0.50}_{-0.47}$ & $16.67^{+3.58}_3.14$ & $10.33^{+2.28}_{-2.13}$ & $0.19^{+0.14}_{-0.12}$ \\

hdfn\_354 & IRX & 189.04882 & 62.17086 & 1.75$^p$ & $<0.30$ & $17.13^{+2.29}_{-1.95}$ & $<0.52$ & $0.82^{+0.10}_{-0.07}$ \\

hdfn\_364 & IRX & 189.30019 & 62.22381 & 2.00$^s$ & $0.22^{+0.13}_{-0.12}$ & $6.83^{+1.35}_{-1.24}$ & $0.57^{+0.35}_{-0.31}$ & $0.63^{+0.14}_{-0.11}$ \\

hdfn\_371 & IRX & 189.18504 & 62.24811 & 1.49$^s$ & $<0.11$ & $1.74^{+0.78}_{-0.68}$ & $<0.11$ & $0.53^{+0.28}_{-0.18}$ \\

hdfn\_401 & IRX & 189.20154 & 62.2379 & 2.00$^s$ & $0.24^{+0.11}_{-0.09}$ & $<0.50$ & $0.64^{+0.30}_{-0.24}$ & $-0.50^{+0.22}_{-0.40}$ \\

hdfn\_476 & IRX & 189.16426 & 62.16842 & 2.35$^s$ & $0.52^{+0.18}_{-0.15}$ & $<0.72$ & $2.32^{+0.79}_{-0.68}$ & $-0.65^{+0.10}_{-0.35}$ \\

hdfn\_490 & IRX & 189.25484 & 62.17116 & 2.13$^p$ & $0.38^{+0.16}_{-0.14}$ & $<0.71$ & $1.23^{+0.53}_{-0.46}$ & $-0.60^{+0.12}_{-0.40}$ \\

%\hline
%\multicolumn{9}{c}{Non-Excess} \\
%\hline

hdfn\_044 & IRNX & 189.06021 & 62.17937 & 1.62$^p$ & $1.90^{+0.31}_{-0.29}$ & $27.89^{+2.62}_{-2.27}$ & $2.64^{+0.43}_{-0.40}$ & $0.49^{+0.07}_{-0.06}$ \\

hdfn\_062 & IRNX & 189.02426 & 62.19673 & 1.48$^s$ & $5.45^{+0.57}_{-0.48}$ & $10.77^{+2.04}_{-1.90}$ & $5.81^{+0.60}_{-0.51}$ & $-0.40^{+0.08}_{-0.09}$ \\

hdfn\_087 & IRNX & 189.09401 & 62.21847 & 2.98$^s$ & $0.61^{+0.18}_{-0.16}$ & $1.59^{+0.88}_{-0.85}$ & $5.70^{+1.72}_{-1.47}$ & $-0.19^{0.22}_{-0.23}$ \\

hdfn\_103 & IRNX & 189.08754 & 62.23701 & 1.77$^p$ & $3.31^{+0.36}_{-0.34}$ & $10.63^{+1.55}_{-1.32}$ & $6.06^{+0.66}_{-0.62}$ & $-0.19^{+0.09}_{-0.08}$ \\

hdfn\_232 & IRNX & 189.2976 & 62.22523 & 2.00$^p$ & $0.73^{+0.19}_{-0.16}$ & $5.09^{+1.18}_{-1.07}$ & $1.95^{+0.50}_{-0.43}$ & $0.17^{+0.16}_{-0.13}$ \\

hdfn\_355 & IRNX & 189.02883 & 62.17255 & 2.51$^s$ & $0.45^{+0.24}_{-0.24}$ & $5.73^{+1.87}_{-1.78}$ & $2.45^{+1.28}_{-1.29}$ & $0.35^{+0.22}_{-0.19}$ \\

hdfn\_356 & IRNX & 189.24513 & 62.17294 & 1.54$^p$ & $0.19^{+0.12}_{-0.12}$ & $12.43^{+1.67}_{-1.56}$ & $0.23^{+0.14}_{-0.14}$ & $0.78^{+0.09}_{-0.07}$ \\

hdfn\_424 & IRNX & 189.3924 & 62.27346 & 1.49$^p$ & $0.75^{+0.31}_{-0.28}$ & $8.48^{+2.21}_{-2.07}$ & $0.80^{+0.33}_{-0.30}$ & $0.34^{+0.19}_{-0.15}$ \\

hdfn\_478 & IRNX & 189.18971 & 62.19094 & 1.53$^s$ & $0.27^{+0.13}_{-0.11}$ & $<0.51$ & $0.32^{+0.16}_{-0.13}$ & $-0.63^{+0.10}_{-0.37}$ \\

hdfn\_045 & pBzK & 189.12127 & 62.17952 & 1.01$^s$ & $1.82^{+0.28}_{-0.26}$ & $20.09^{+2.11}_{-1.78}$ & $0.61^{+0.09}_{-0.09}$ & $0.38^{+0.06}_{-0.08}$ \\

\hline
\end{tabular}
\end{table*}

\pagebreak

\section{Results}
\label{sec:results}

\subsection{Direct Detections}
\label{sub:doubleDs}

We match the sBzK galaxies in 4Ms CDFS and 2Ms CDFN to X-ray point sources using a 1.25 arcsecond search radius to establish direct detections. The BzK source density is sufficiently low that at 1.25 arcseconds all matches are unique, and the astrometry corrections detailed in Section \ref{sub:xray_redux} have made the overall offsets between the optical and X-ray positions relatively small. We directly detect \textbf{5.8$\pm$0.8\% (49/851)} of sBzKs in the 4Ms CDFS image and \textbf{5.4$\pm$1.0\% (32/594)} of sBzKs in the 2Ms CDFN image (see Tables 2 and 3). The rest-frame 2-10 keV luminosities ($L_{2-10 keV}$; derived from $F_{0.5-2 keV}$) of sBzKs range from $1.1\times10^{41}-3.5\times10^{44}$ erg s$^{-1}$ with median $L_{2-10 keV}=3.5\times10^{42}$ erg s$^{-1}$. Approximately 70\% (57/81) of the detected sBzKs possess $L_{2-10 keV}>10^{42}$ erg s$^{-1}$ and it is reasonable to assume all of these objects host AGN. The X-ray emission of the remaining 26 sBzKs is probably also mostly from AGN, but a significant contribution from stellar processes is also possible for these lower luminosity X-ray sources. 

\begin{figure}
\includegraphics[width=80mm]{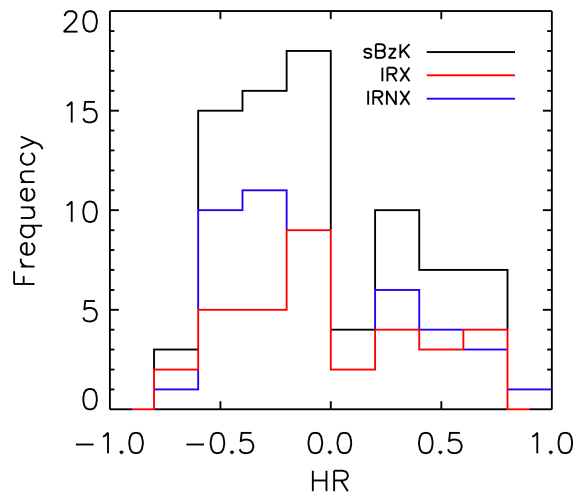}

\caption{Histogram of HR of directly detected sBzK galaxies. The black line represents all sBzKs, the red line represents IRXs and the blue line represents IRNXs.}
\label{fig:hr_hist}
\end{figure}

The detection rate of X-ray sources is a strong function of $L_{IR}$ (see Figure \ref{fig:xs_vs_lir}). This agrees with previous observations that AGN are more prevalent amongst the most IR luminous galaxies (\citealt{sanders96}; \citealt{risaliti00}). The HR distribution of X-ray detected sBzK galaxies is notionally bimodal (Figure \ref{fig:hr_hist}), covering a wide range of HR (-0.70 to 0.82). Using a simple photo-electric absorption model, an AGN at $z\sim2$ with power-law emission $\Gamma=1.9$ obscured by material with N$_H>10^{23}$ cm$^{-2}$ should possess a HR $>-0.08$. Therefore 44\% (36/81) of the X-ray detected sBzKs possess N$_H=10^{23}$ cm$^{-2}$ or greater. This suggests the sBzKs are a mixture of obscured AGN, unobscured AGN and purely star-forming galaxies.

Splitting the sBzK galaxies into IRX and IRNX sub-samples we directly detect 13.2$\pm$1.9\% (47/356) and 3.9$\pm$0.7\% (34/879) respectively. It is clear that a greater proportion of IRXs than IRNXs are detected in X-rays. The IRXs are also significantly more X-ray luminous (median $L_{2-10 keV}=6.44\times10^{42}$ erg s$^{-1}$) than IRNXs (median $L_{2-10 keV}=2.26\times10^{42}$ erg s$^{-1}$). On the other hand, the higher detection fraction of IR Excess sBzKs may be more related to their IR luminosity, rather than the IR excess per se. Fig.~4 clearly demonstrates a very strong correlation between the excess parameter and $L_{\rm IR}$, such that almost all of the most IR-luminous BzK galaxies also exhibit an IR excess. 

\begin{figure}
\includegraphics[width=80mm]{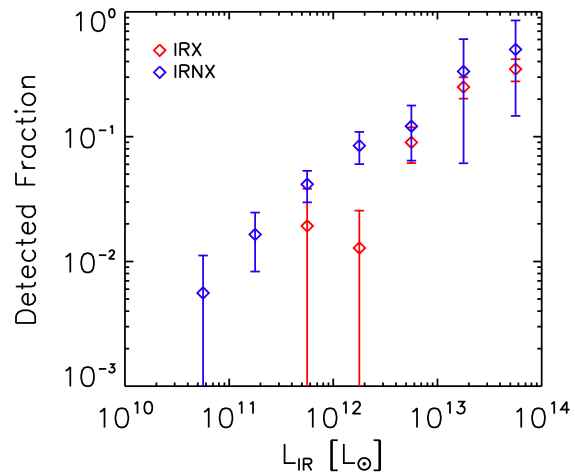}

\caption{Plot of detected fraction of sBzKs against $L_{IR}$ where red diamonds are IRXs, blue diamonds are IRNXs and black diamonds are all sBzKs. Multinomial errors for sBzK detected fraction calculated using relationship of \citet{nandra07}.}
\label{fig:detect_frac}
\end{figure}

If we bin the sBzKs by IR luminosity and then compare IRXs and IRNXs, as in Figure \ref{fig:detect_frac}, the detected fractions are largely similar. If anything, when controlled for IR luminosity, the IRNXs show a higher detected fraction, although this is not statistically significant. A similar fraction of IRXs and IRNXs possess $L_{2-10 keV}>10^{42}$ erg s$^{-1}$ (74\% and 65\% respectively), so the conclusion should also hold for the AGN fraction as well (i.e. as opposed to the detected fraction), especially when considering that the strongest star-formation will be in the most IR-luminous (and hence excess) galaxies.

The IRXs and IRNXs have very similar distributions of HR (see Figure \ref{fig:hr_hist}). There are 14 (30\%) IRXs and 11 (32\%) IRNXs with HR $>0.2$. These hard X-ray sources are almost certainly obscured AGN candidates, but there is no evidence that there is a heavier concentration of heavily obscured AGN within the IRX sample compared to the IRNX sample. A K-S test shows that, based on their HR distributions, the IRXs and IRNXs are consistent with each other (null hypothesis probability p=0.829). There are 4 sBzKs in our sample in CDFS that have previously been confirmed using X-ray spectroscopy as containing Compton thick AGN: cdfs4Ms\_077 and cdfs4Ms\_283 by \citet{feruglio11} and cdfs4Ms\_126 and cdfs4Ms\_226 by \citet{brightman12b} (note that cdfs4Ms\_077 was also identified by \citealt{brightman12b} as containing a Compton thick AGN). Only cdfs4Ms\_226 is an IRNX\textbf{, while cdfs4Ms\_077, cdfs4Ms\_126 and cdfs4Ms\_283 are IRXs. The fraction of IRXs confirmed as Compton thick AGN from their X-ray spectra is lower than expected (3/49), but the X-ray detected IRX sample is too small to draw any strong conclusions regarding the nature of IRXs as a whole.}

\subsection{Stacking sBzKs}
\label{sub:xsstack}

In addition to analyzing the direct detections we also stacked the X-ray undetected sBzKs in 4Ms CDFS and 2Ms CDFN to probe their average X-ray properties. In the first instance the X-ray photons are stacked in 3 broad bands: $0.5-2$ keV (soft), $2-7$ keV (hard) and $0.5-7$ keV (full). The stacked photon count rates are then converted to fluxes assuming a $\Gamma=1.4$ power-law: $F_{0.5-2keV}$ (soft), $F_{2-10keV}$ (hard), and $F_{0.5-10keV}$ (full). Starting with the full sample, we stacked 370 sBzKs in 4Ms CDFS and find a highly significant signal in the soft and hard band. The stacked signal possessed a HR~$=-0.53\pm0.08$. We stacked 318 sBzKs in 2Ms CDFN and again found significant soft and hard band signals, with a HR~$=-0.47\pm0.09$. The stacked flux of the X-ray undetected sBzKs in CDFN is higher than for those in CDFS (see Table 5) because the 4Ms CDFS is deeper, and the brightest sources stacked in the CDFN 2Ms image would be detected and resolved out in the CDFS 4Ms image. For the total sample, both CDFS and CDFN sBzK galaxies exhibit soft HRs in these broad bands and there is no evidence to suggest a substantial proportion of the sBzKs contain heavily obscured AGN.

Splitting the sBzKs using the IR excess criterion (Equation \ref{eq:xs_crit}), we stacked X-ray undetected 72 IRXs and 298 IRNXs in 4Ms CDFS. The stacked signal of the X-ray undetected IRXs is significantly detected in both soft and hard band with a HR~$=-0.48\pm0.11$. The stacked signal of X-ray undetected IRNXs is also significantly detected in the soft and hard band, but is more X-ray faint (see Table 5) and has a softer HR ($-0.55\pm0.11$). The higher flux of the stacked IRXs is in agreement with our observations of directly detected sources (Section \ref{sub:doubleDs}). Again, however, this can be attributed to the higher IR luminosity for the excess sources, whose median  $L_{\rm IR}=3.6 \times 10^{12} L_{\odot}$ is over an order of magnitude higher than the IRNX sources (median  $L_{\rm IR}=3.1 \times 10^{11} L_{\odot}$).  Additionally the HR and effective $\Gamma$ of X-ray undetected IRXs and IRNXs in CDFS are consistent with each other at the $1\sigma$ level (see Table~5) which suggests the composition of the sub-samples is similar.

Turning to the 2Ms CDFN we stacked 107 X-ray undetected IRXs and 211 X-ray undetected IRNXs. As with the stacking in 4Ms CDFS, the IRXs are significantly detected in both the soft and hard bands. The HR ($-0.50\pm0.12$), agrees with the stacked signal of 4Ms CDFS X-ray undetected IRXs at the $1\sigma$ level. The X-ray undetected IRNXs are significantly detected in the soft band (S/N = 8.2) but the hard band is a low significance detection (S/N = 2.7). The stacked fluxes are again higher than 4Ms CDFS X-ray undetected IRNXs (see Table 5) and the HR~$=-0.45\pm0.15$. The HRs of X-ray undetected IRXs and IRNXs are consistent at the 1$\sigma$ level although the low significance hard band detection of X-ray undetected IRNXs means the HR is probably softer than our current estimate. Ultimately all stacks of X-ray undetected sBzKs and the IRX and IRNX subsamples agree at the $1\sigma$ level (Table \ref{tab:xsstack}) and there is no evidence from broad band stacking that heavily obscured AGN are more prevalent in IRXs than IRNXs.

As discussed in Section \ref{sub:sample}, the IR excess criterion (Equation \ref{eq:xs_crit}) is positively correlated with redshift and $L_{IR}$ (Figures \ref{fig:xs_vs_z} and \ref{fig:xs_vs_lir} respectively) and this is likely to account for the higher X-ray fluxes of the IRXs. To investigate this further, we have stacked samples binned by redshift and $L_{IR}$. The sBzK galaxy samples from CDFS and CDFN have been combined to maximise the S/N ratio in each redshift and $L_{IR}$ bin.

The X-ray undetected sBzKs are split into 3 redshift bins with ranges that have been chosen to distribute the sources as evenly as possible: $1.2\le z<1.8$, $1.8\le z<2.4$ and $2.4\le z<3.0$. The stacked emission of X-ray undetected sBzKs with $1.2\le z<1.8$ and $1.8\le z<2.4$ possess HR~$=-0.47\pm0.10$ and HR$=-0.48\pm0.09$ respectively and exhibit similar X-ray fluxes (see Table 6). The effective $\Gamma$ values are consistent with those observed for X-ray undetected IRX and IRNX subsamples at the $1\sigma$ level. The $2.4\le z<3.0$ X-ray undetected sBzKs have a similar flux to the lower redshift sources in the soft band, but no detection in the hard band (S/N = 0.6). The HR is therefore very uncertain, but inferring the hard flux from the full band detection shows X-ray colours which are entirely consistent with the lower redshift bins. As the fluxes of each redshift-binned subsample are all similar, this naturally implies that the higher redshift sources are on average more luminous in the X-ray, but this is likely to be a selection effect due to the fact that the parent sample is magnitude limited, meaning that the higher redshift sources need to be more luminous overall to enter the sample. 

The X-ray undetected sBzKs were also split into 4 $L_{IR}$ bins to explore the connection between $L_{IR}$ and the IR excess criterion (Section \ref{sec:data}): $L_{IR}<10^{11} L_{\odot}$, $10^{11} L_{\odot}\le L_{IR}<10^{12} L_{\odot}$ , $10^{12} L_{\odot}\le L_{IR}<10^{13} L_{\odot}$ and $L_{IR}\ge10^{13} L_{\odot}$. There is a strong correlation between X-ray brightness and $L_{IR}$ of the sBzKs (see Table 6). This result is in agreement with our observations of directly detected X-ray sources. All bins are significantly detected in hard and soft bands except for $L_{IR}<10^{11} L_{\odot}$  which has a low significance detection in the hard band (S/N=2.1). The HR values of all the $L_{IR}$ binned sBzK sub-samples are consistent with one another at the $1\sigma$ level and the stacked signal of IRXs and IRNXs in CDFN and CDFS. As with the redshift selected galaxies, none of the $L_{IR}$ bins appear to preferentially select heavily obscured AGN, but the BzK galaxies which are more luminous in the IR are brighter in the X-ray. 

\subsection{Stacking Smaller Bands}
\label{sub: bandstack}

\begin{table*}
\begin{minipage}{175mm}
\begin{center}
\caption{Narrow band stacking of X-ray undetected sBzK galaxies in 4Ms CDFS and 2Ms CDFN. Column (1): energy range of narrow band; column (2): number of sBzKs stacked; column (3): S/N of stacked sBzKs; column (4): narrow band flux of stacked sBzKs, units $10^{-18}$ erg cm$^{-2}$ s$^{-1}$; column (5): number of IRXs stacked; column (6): S/N of stacked IRXs; column (7): narrow band flux of stacked IRXs, units $10^{-18}$ erg cm$^{-2}$ s$^{-1}$; column (8): number of IRNXs stacked; column (9): S/N of stacked IRNXs; column (10): narrow band flux of stacked IRNXs, units $10^{-18}$ erg cm$^{-2}$ s$^{-1}$.}
\begin{tabular}{@{} l c c c c c c c c c c c c }
\hline
& & \multicolumn{3}{c}{sBzKs} & & \multicolumn{3}{c}{IRXs} & & \multicolumn{3}{c}{IRNXs}  \\
Band (keV) & & No. Src & S/N & Flux & & No. Src & S/N & Flux & & No. Src & S/N & Flux \\
(1) & & (2) & (3) & (4) & & (5) & (6) & (7) & & (8) & (9) & (10)\\
\hline
0.5-1.0 & & 688 & 10.15 & $40.63\pm3.94$ & & 179 & 6.51 & $51.35\pm8.08$ & & 509 & 7.93 & $36.87\pm4.50$ \\

1.0-1.5 & & 688 & 15.21 & $8.07\pm0.55$ & & 179 & 11.71 & $14.83\pm1.30$ & & 509 & 10.47 & $5.70\pm0.59$ \\

1.5-2.0 & & 688 & 9.27 & $1.18\pm0.13$ & & 179 & 6.22 & $1.65\pm0.28$ & & 509 & 7.09 & $1.01\pm0.15$ \\

2.0-2.5 & & 688 & 3.92 & $1.12\pm0.29$ & & 179 & 3.12 & $1.84\pm0.60$ & & 509 & 2.72 & $0.86\pm0.32$ \\

2.5-3.5 & & 688 & 3.44 & $5.34\pm1.62$ & & 179 & 1.32 & $2.86\pm3.16$ & & 509 & 3.19 & $6.21\pm1.89$ \\

3.5-5.0 & & 688 & 2.40 & $5.01\pm2.09$ & & 179 & 2.03 & $9.92\pm4.36$ & & 509 & 1.61 & $3.29\pm2.38$ \\

5.0-7.0 & & 688 & 1.57 & $6.92\pm4.23$ & & 179 & 2.25 & $17.29\pm8.76$ & & 509 & 0.51 & $3.28\pm4.82$ \\
\hline
\end{tabular}
\end{center}
\end{minipage}
\label{tab:narrow_bands}
\end{table*}

\begin{figure}
\includegraphics[width=80mm]{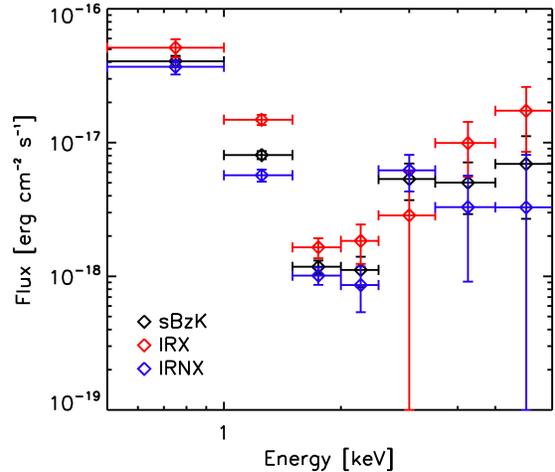}

\caption{Flux estimates for narrow band stacking of IRXs (red diamonds) and IRNXs (blue diamonds) galaxies in the 4Ms CDFS and 2Ms CDFN. The horizontal error bars represent the width of the energy band that is being stacked. The IRXs stacked in the two highest energy bands are not significantly detected.}
\label{fig:cdfs_narrow_flux}
\end{figure}

When stacked in broad bands (Section \ref{sub:xsstack}) the X-ray undetected IRXs and IRNXs both appear to have relatively soft X-ray spectra, with effective $\Gamma\sim2$. Based on this analysis there is no evidence for obscured AGN in either subsample. The broad bands, however, may mask more complex spectral features and give a misleading picture of the actual spectral shape and hence the level of obscuration (see e.g. \citealt{brightman12a}). We have therefore stacked the sBzKs in narrower energy bands to obtain a more accurate representation of their spectra (Table 4 and Figure \ref{fig:cdfs_narrow_flux}). The bands in Table 4 have been designed to provide the optimum balance of resolution and signal strength. Narrower bands have fewer counts so the CDFS and CDFN sBzKs have been combined to maximise the S/N ratio in these bands. This analysis shows clearly that the naive interpretation of the broad band colours gives a misleading impression of the true spectral shape, which does not resemble a single power law. The total spectrum is clearly "composite" in nature with a soft component dominating the soft energies, and a flat, hard component coming through at high energies. The latter is a clear signature of obscured AGN in the sBzK galaxies. The soft component is more ambiguous and could originate from a range of sources. There could be contributions from unobscured AGN, scattered light in heavily obscured AGN and/or X-rays from pure star forming processes. The stacked spectrum indicates that there is a mixture of these effects within the sample. 

Splitting the sample into IRXs and IRNXs, we observe that they have remarkably similar spectral shape with the only noticeable difference being that IRNXs are consistently less X-ray bright (see Figure \ref{fig:cdfs_narrow_flux}), just as was observed for broad band stacking (see Section \ref{sub:xsstack}). Thus, while the narrow band analysis confirms the presence of residual AGN activity in the population of sBzK galaxies as a whole, there is no evidence that obscured AGN or AGN generally are preferentially hosted by those with excess emission at 24 $\mu$m.

\subsection{Passive galaxies}
\label{sub: passive}
There are 21 passively evolving BzK's (pBzK) in CDFS and 5 pBzK's in CDFN. We directly detect 3 pBzK's in 4Ms CDFS and 1 pBzK in 2Ms CDFN, see Tables 2 and 3. While these numbers are small, and hence the uncertainties large, the pBzK detection fraction (15$\pm$8.0\%; 4/26) is actually greater than for sBzKs (7$\pm$0.7\%; 81/1235). The sources cdfs4Ms\_301, cdfs4Ms\_457 and hdfn\_045 have hard signals which suggest the presence of obscured AGN.  The source cdfs4Ms\_318 lacks a hard band detection but $L_{0.5-2.0keV}=1.1\times10^{42}$ erg s$^{-1}$ (derived from $F_{0.5-2kev}$). The X-ray emission probably originates from either unobscured AGN activity or from a hot gas halo within the galaxy \citep{forman85,sarazin97}. Intense star-formation as the origin of the observed X-ray emission is extremely unlikely, recalling that these objects have been selected as passively evolving galaxies. Stacking the X-ray emission of the 13 undetected pBzKs in CDFS and CDFN yields a marginal signal in the soft band, with average $F_{0.5-2keV}=2.23\pm0.62\times10^{-17}$ erg cm$^{-2}$ s$^{-1}$. Stacking the hard band flux does not result in a significant detection. 

\section{Comparison to previous results}
\label{sec:comparison}

\citet{daddi07b} stacked the X-ray emission of 59 IRXs and 175 IRNXs that were not directly detected in X-rays using a 1Ms CDFS image and source list. Their sample of IRXs had a much harder stacked signal (effective $\Gamma=0.8^{+0.4}_{-0.3}$) than our stacked sample in the 4Ms CDFS and 2Ms CDFN data (see Table \ref{tab:xsstack}). There are a number of possible reasons for this discrepancy, including differences in the parent BzK sample and deeper X-ray data in our case. To help understand this discrepancy we stacked our samples of IRXs and IRNXs using only the 1Ms CDFS images and source list to provide a direct comparison (Table \ref{tab:detstack}). Due to the shallower data, fewer sBzKs are directly detected in the 1Ms CDFS so we stack more X-ray undetected IRXs (80) and IRNXs (311). Our samples of X-ray undetected IRXs and IRNXs are different to (and larger than) those of \citet{daddi07b} because of the updated photometry of GOODS-MUSIC v2 including deeper 24$\mu$m data used in our work (\citealt{desantis07}). 

The X-ray undetected IRXs were significantly detected in both the soft (S/N = 5.6) and hard (S/N = 3.6) bands with HR $=-0.16\pm0.16$ and corresponding effective $\Gamma=1.3^{+0.4}_{-0.3}$. This  effective $\Gamma$ is clearly rather harder then we obtained earlier, and while formally softer than that derived by  \citet{daddi07b} it is consistent with their results of at the 1$\sigma$ level.  The X-ray undetected IRNXs are significantly detected in the soft band (S/N = 7.6), but the hard band is a low significance detection (S/N = 2.2), something which \citet{daddi07b} also found. The HR $=-0.52\pm0.17$ and corresponding effective $\Gamma=2.1^{+0.6}_{-0.4}$ are poorly constrained but are also consistent with the effective $\Gamma\sim1.7$ of \citet{daddi07b} at the $1\sigma$ level.

\begin{figure}
\includegraphics[width=80mm]{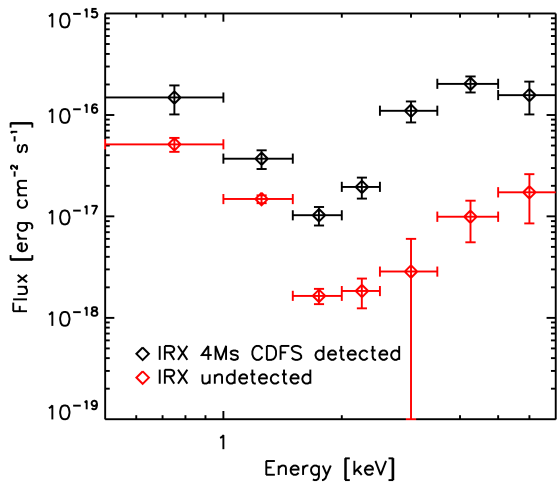}

\caption{Flux estimates for narrow band stacking of 72 X-ray undetected IRXs (red diamonds) and 6 IRXs that were only detected in the 4Ms CDFS data (black diamonds) using 4Ms CDFS images. The horizontal error bars represent the width of the energy band that is being stacked.}
\label{fig:cdfs_detstack_flux}
\end{figure}

Our analysis, and that of \citet{daddi07b}, shows that the spectrum is more complex than a simple power, so a naive comparison of the two spectral indices does not give the full picture. Nonetheless it is clear that we find that the overall spectrum of X-ray undetected IRXs is softer. The above analysis shows that this can be partially, but not completely explained by the difference in the parent BzK sample. A further possibility is that the 1Ms stack is contaminated by hard sources undetected in the 1Ms image, but detected and resolved out in the 4Ms data. There are 22 sBzKs (9 IRXs; 13 IRNXs) which are detected in the 4Ms CDFS image but are undetected (i.e. fall below the detection limit) in the 1Ms CDFS image (see Table \ref{tab:cdfs_direct}). Of these 22 sBzKs, 6 IRXs and 6 IRNXs were stacked in the 1Ms CDFS image. The 3 IRXs and 6 IRNXs that were directly detected in the 4Ms CDFS images were omitted from the stacking as they fell outside the 5.5 arcminute inclusion radius (see Section \ref{sub:stacking}). Focusing on the X-ray detected IRXs (because their signal softened more dramatically) the average HR $=0.14\pm0.09$, with effective $\Gamma=0.7^{+0.2}_{-0.1}$ much harder than the stacked signal of IRXs undetected in the 4Ms image (HR $=-0.48\pm0.11$). This strongly suggests that the 1Ms CDFS stacked signal is indeed contaminated by hard sources which are now detected in the 4Ms CDFS data. Narrow band stacking of IRXs that are detected in the 4Ms CDFS but not the 1Ms CDFS (Figure \ref{fig:cdfs_detstack_flux}) better illustrates the full effect of the 4Ms detected population. The stacked spectral shape of X-ray undetected IRXs and 4Ms detected IRXs is similar up to 1.5 keV (approximately 4.5 keV rest-frame). At higher energies the stacked signals diverge and the 4Ms detected IRXs dominate the X-ray undetected IRXs. Including the 4Ms detected sBzKs in the stacking analysis (as in the 1Ms CDFS) causes the X-ray undetected IRXs to appear to have a much harder signal. We conclude that this is the main reason why the stacked signal of X-ray undetected IRXs in \citet{daddi07b} is harder than the stacked signal we find for the 4Ms CDFS X-ray undetected IRXs.

\citet{alexander11} took a more in depth look at the X-ray properties of sBzKs, limiting their sample of sBzKs (an updated version of the \citet{daddi07b} sample) to those with $<5.5$ arcminute OAA and cross-correlating them with the 4Ms CDFS catalogue of \citet{xue11} using a 1.5 arcsecond matching radius. They directly detect 47 sBzKs, where 60\% are IRXs and the remaining 40\% are IRNXs. Additionally 23\% (11/47) of the sBzKs are classified as heavily obscured AGN because they exhibit a hard X-ray spectral shape ($\Gamma\le1$). If we apply a similar cut in OAA to our sample (see Table 2) we are left with 35 X-ray detected sBzKs, with 51\% classified as IRXs and 49\% classified as IRNXs. Using the HR of these sBzKs we define 46\% of these sBzKs as heavily obscured (HR $>-0.08$; see Section \ref{sub:doubleDs}).

We cross correlate the \citet{alexander11} X-ray detected sBzKs with the 35 sBzKs with OAA $<5.5$ in our sample (Table 2) using a search radius of 1.25 arcseconds and find that 23 sBzKs are common to both samples. There 19 IRXs and 14 IRNXs within this ``shared" sample. There are 8 sources which are defined as heavily obscured by \citet{alexander11} and 9 sources which we define as heavily obscured in this work (see Section \ref{sub:doubleDs}). These heavily obscured sources show good agreement, with 7 of the sBzKs defined as heavily obscured in both this work and \citet{alexander11}. The X-ray fluxes from the \citet{xue11} 4Ms CDFS catalog used in the \citet{alexander11} were calculated by extracting X-ray data over 0.5-2 keV (soft), 2-8 keV (hard) and 0.5-8 keV (full) ranges sample and then integrating over power-laws individually derived from the hard-to-soft band ratio ($H/S$) of each source. We recalculated fluxes for the 23 ``shared" sources using the same energy ranges as the \citet{xue11} 4Ms CDFS catalogue and $\Gamma=1.4$. There are inevitably some discrepancies between the fluxes presented in this work and those presented by \citet{xue11} because of the difference in $\Gamma$ values used to calculate flux. Irrespective of this the majority of fluxes show good agreement (lie within 1$\sigma$ error bars).

The \citet{alexander11} sample has 24 X-ray detected sBzKs which were not present in our sample, 8 of which we identified as sBzKs but were not detected in our 4Ms CDFS catalogue, while the remaining 16 were not in our sBzK sample. The \citet{xue11} 4Ms CDFS catalogue has 740 sources while our more stringent 4Ms catalogue has 569 sources \textbf{(see Section \ref{sub:xray_redux})}, so it is to be expected that they detect a larger number of sBzKs. The 8 galaxies we identify as sBzKs but do not detect in our catalogue are all low significance detections (have low binomial detection probabilities) in the \citet{xue11} 4Ms CDFS catalogue, hence likely lie just below the detection threshold of our catalogue. The remaining 16 X-ray detected sBzKs in the \citet{alexander11} sample were omitted from our sBzK sample because they did not satisfy one or more of the criteria given in Section \ref{sub:optir}. We have tried to keep our sample as similar as possible to that of \citet{daddi07b} but there are several key differences. We use a different IRAC $3.6\mu$m selected catalogue to ensure the 24$\mu$m sources are deblended. Additionally 24$\mu$m source photometry in our sample was extracted using the ConvPhot software \citep{desantis07} whereas the \citet{daddi07b} sample uses a more basic fixed matching radius method. These are the likely source of the discrepancies between our sBzK selection. There are also 12 X-ray detected sBzKs in our sample which are not present in the \citet{alexander11} sample. Again this is due to the aforementioned discrepancies between our respective X-ray point source catalogues and optical sample selection criteria.

\citet{alexander11} stack the remaining X-ray undetected sBzKs in a similar fashion to \citet{daddi07b} using the 4Ms CDFS images. They stack 47 X-ray undetected IRXs and 116 X-ray undetected IRNXs and find effective $\Gamma=1.4\pm0.3$ and $\Gamma=2.0\pm0.4$ respectively. The effective $\Gamma$ values of the stacked emission of X-ray undetected IRXs and IRNXs agree with those we calculate using our samples in 4Ms CDFS at the 1 $\sigma$ level (see Table 5). The effective $\Gamma$ of the X-ray undetected IRXs is also significantly softer than the value calculated in \citet{daddi07b}. Similar to us, \citet{alexander11} conclude that the majority of hard sources have been resolved in the 4Ms CDFS data and that there is no longer strong evidence that heavily obscured AGN are the dominant population amongst the remaining X-ray undetected sBzKs.

\section{Discussion}
\label{sec:discussion}

Using the deepest data available, we have investigated the X-ray properties of a large sample of star-forming BzK-selected galaxies in the redshift range $z=1.2-3.0$ in the CDFN and CDFS. The X-ray emission is particularly important when assessing the AGN content of these galaxies. Specifically, a direct detection of the X-ray emission makes the presence of an actively growing supermassive black hole very likely, as the implied luminosity (L$_X>10^{42}$ erg s$^{-1}$) can only be exceeded by the most extreme star-forming galaxies. We find a relatively high \textbf{(6~\% approximately)}, direct X-ray detection fraction for the sBzK galaxies overall, indicating a very substantial AGN content. Also, consistent with previous analysis, we find a very large fraction of the directly-detected sBzK to be heavily obscured, based on their hardness ratios. The implied column densities are substantial (usually N$_H>10^{23}$~cm$^{-2}$). Four of the X-ray detected sBzKs have been identified as Compton thick AGN based on their X-ray spectra in previous work: cdfs4Ms\_077 and cdfs4Ms\_283 by \citet{feruglio11} and cdfs4Ms\_126 and cdfs4Ms\_226 by \citet{brightman12b}. \textbf{Of these four Compton thick AGN, 3 are IRXs and 1 is an IRNX.}

Stacking the emission from the sBzKs which are not detected directly shows a strong signal in both the soft and hard X-ray bands. By stacking in relatively narrow energy bands we have constructed a crude spectrum of these undetected objects. This is composite in nature, with a soft component dominating at low energies and harder emission at high energy. The hard spectrum above $\sim 2$~keV is a clear signature of obscured AGN activity in part of the sBzK population at a level below the threshold of the direct detection. Part of the soft emission could be scattered light from these obscured AGN. On the other hand it is also likely that unobscured, low luminosity unobscured AGN emission and soft X-rays from star-forming processes also contributes to the stack at low energies. We are likely therefore seeing a mixture of obscured AGN, unobscured AGN and star-forming galaxies. Disentangling these is rather difficult. On the whole, however, our results support the idea that sBzK galaxies contain a rather active population of obscured AGN, confirming previous work \citep{daddi07b,alexander11}. 

Where our results diverge somewhat from previous analysis (e.g. that of \citealt{daddi07b}) is when considering subsamples of the sBzK population. Specifically, we have examined the direct detections and stacked X-rays from subsamples based on their ``IR Excess", the ratio of the total star formation rate to that of the dust-corrected UV SFR. IRXs do exhibit a significantly higher X-ray detection fraction (by a factor $3-4$) than IRNXs. However, the majority of this can be relatively simply explained given the extremely strong correlation between the IR excess parameter and the total infrared luminosity. When controlling for IR luminosity, there is no statistically-significant difference in the X-ray detected fraction comparing IRXs and IRNXs.

Stacking results for the fainter sources confirm this effect. The stacked IRXs give a slightly higher flux, but the hardness ratios of the IRXs and IRNXs are very similar. Stacking in narrow bands also shows that the two populations have very similar spectra to each other. In particular, both stacks show a hard component indicating the presence of obscured AGN in some of the objects, too faint to be detected individually. Overall, it seems clear that the IR excess parameter is not decisive in selecting obscured AGN or AGN more generally from the sBzK population. The fact that the IR-luminous sBzKs show a high X-ray detection fraction could be attributable to a number of factors. First the BzK selection will tend to yield samples of galaxies with high stellar mass and red colours, which are known to preferentially host luminous AGN (e.g. \citealt{nandra07}; \citealt{bundy08}). Galaxies with high IR luminosities will also tend to be the most luminous and massive sources themselves. The high IR luminosities and star formation rates also suggest copious amounts of gas and dust, which could account for the high levels of obscuration in these sources. On the other hand, it is puzzling why the presence of an AGN as inferred by the presence of (hard) X-ray emission apparently bears so little relation to excess 24$\mu$m emission. Apparently there are many galaxies without a strong hot dust component which nonetheless must host an AGN based on their X-ray emission and, similarly, many which have very strong 24$\mu$m but without a powerful AGN. 

\citet{daddi07b} have suggested that the obscuration levels in the IRXs may in fact be so high that they are ``Compton thick" AGN, where even the X-ray emission is suppressed and observed only in scattered light. If so a considerable amount of additional black hole growth may be hiding in these objects. This issue has been investigated further by \citealt{alexander11} who examined the X-ray spectral properties of directly detected sBzKs using the 4Ms CDFS data. Their analysis resulted in a sharp downward revision in the estimated Compton thick AGN fraction in BzKs compared to \citet{daddi07b}, who used X-ray stacking analysis to show that the hard X-ray emission of the IRXs resembled the typical spectrum of Compton thick AGN, with a very hard spectrum at high energies. 

Our stacking results confirm the presence of a hard high energy spectrum in the sBzK population. Overall the spectrum of the stacked sources is of a ``composite" nature, with soft emission either from star formation or unobscured AGN, and a hard signal from obscured AGN.  Determining whether the obscuration in the obscured AGN is Compton thin or Compton thick is extremely difficult, as the X-ray continuum from moderately obscured Compton thin AGN ($N_{\rm H}~\sim 10^{23}$ cm$^{-2}$) and the reflection dominated spectrum of  Compton thick AGN is quite similar. The telltale signature is the intense iron K$\alpha$ line (E = 6.4keV rest-frame) predicted in the Compton thick case. This signature has been found in at least one sBzK galaxy \citep{feruglio11} and is inferred by \citet{alexander11} to be present in a handful of additional sBzKs, based on the co-addition of the X-ray spectra of several directly detected sBzKs with hard X-ray colours in the 4Ms CDFS data. As discussed above the conclusion of widespread Compton thick AGN in BzKs by \citet{daddi07b} was based on the shape of the stacked spectrum. With a much larger sample of BzK galaxies, and far deeper X-ray data, it may have been hoped that our stacking analysis would reveal the iron K$\alpha$ line predicted in the Compton thick AGN case. In the event, however, our results neither confirm not deny the presence of intense iron K$\alpha$ emission in the galaxies. The primary issue is that the deeper X-ray data are able to resolve out the brighter X-ray sources from the stack, which carry the majority of the signal-to-noise ratio. As noted by \citet{alexander11}, and confirmed by us, a great number of the objects directly detected in the 4Ms CDFS Chandra data, but not in the 1Ms data used by \citet{daddi07b}, have very hard spectra. These sources are undoubtedly obscured, and a fraction of them may be Compton thick, but once removed from the stack evidence for a  Compton thick AGN in the residual sBzK population is weakened, rather than strengthened. We conclude that it is particularly difficult to base estimates of the contribution of Compton thick AGN on a stacking analysis, as stacking will result in a mixture of X-rays from obscured AGN, unobscured AGN and stellar processes, and can be strongly affected or even dominated by a relatively small number of sub-threshold sources. Definitive conclusions regarding the presence of the tell-tale iron K$\alpha$ emission in faint Compton thick AGN awaits future experiments with much larger collecting area than {\it Chandra}, such as the {\it Athena} concept \citep{barcons12}. 

In addition to impacting on our knowledge of the AGN content of sBzK galaxies, our X-ray results also have interesting implications about the nature of the IR-excess sBzK population. The basis of the X-ray analysis of the IRXs is that the additional emission at 24$\mu$m could be due to hot dust emission powered by a buried AGN. Our analysis has clearly shown, however, that the AGN content (and obscuration properties) of IRXs and IRNXs are very similar. This raises the question of the origin of the large discrepancy between the 24$\mu$m-derived star formation rates and the dust-corrected UV measurements. As these are apparently not due to AGN activity it indicates that the derivation of one or other of the SFRs using the methodology adopted in the current paper is faulty. The discrepancy is largest at high IR luminosity, where the 24$\mu$m SFR greatly exceeds the UV estimates, in extreme cases by 2 orders of magnitude. This implies either that the 24$\mu$m SFR is overestimated, the UV underestimated, or both. Recent results from {\it Herschel} suggest that both effects could be in play. Firstly, 24$\mu$m fluxes appear systematically to over-estimate the SFRs, especially of high redshift galaxies \citep{elbaz11,nordon12}. Secondly, at the highest IR luminosities and thus SFRs, the lines of sight to the star forming regions may become completely opaque to UV emission, being obscured by optically thick dust \citep{goldader02}. This makes the rest-frame UV emission, even if corrected for dust extinction, an extremely unreliable estimate of the star formation rate and specifically it will lead one to underestimate the SFR. 

A final interesting point from the current work stems from the X-ray analysis of the passive BzK galaxies. While our sample of pBzK is not large (26 objects) it is sufficient to be able to make the first estimate of the AGN content of these galaxies. Of 18 pBzKs in the redshift range $z=1.2-3.0$ we find two X-ray detections. Expanding out the full redshift range spanned by our BzK sample the detections rise to 4/26 objects. While the number statistics are still small, the X-ray detection rate and AGN content of the pBzKs is at least consistent with and perhaps even slightly higher than detection rate in sBzKs. The high AGN fraction in sBzKs has been interpreted by \citet{daddi07b} as evidence for rapid coeval star formation and black hole growth in this galaxy class. It appears, however, that this is not robust, as BzK galaxies without rapid star formation also show an unusually high fraction of AGN. So while a large amount of black hole growth in galaxies at $z\sim2$ does indeed occur in galaxies that are also actively growing their stellar component, this is arguably more likely to be due to the fact that most massive galaxies are high $z\sim2$ are star forming galaxies. It does not necessarily present a causal and or astrophysical link between black hole growth and star formation.

\section{Summary}
\label{sec:summary}

Using 4Ms CDFS and 2Ms CDFN~\textit{Chandra}/ACIS imaging we have investigated the X-ray properties of BzK-selected galaxies at $z\sim2$. The AGN content and obscuration properties have been studied using a combination of direct detection, broad band stacking and narrow band stacking analyses. The main results are listed below:

\begin{enumerate}
\item We directly detect \textbf{$5.6\pm0.6\%$ (81/1445)} of star-forming sBzKs in X-rays, finding 49 sBzKs in the 4Ms CDFS and 32 sBzKs in the 2Ms CDFN \textit{Chandra} ACIS-I images. The direct detection limits are such that almost all of these will host AGN. We directly detect a greater proportion of IR Excess sBzKs relative to IR Non-Excess sBzKs. However, this can be attributed entirely to the fact that the IR excess criterion also selects the most IR luminous sBzKs. The X-ray detected fractions of IR Excess sBzKs and IR Non-Excess sBzKs are very similar when binned by $L_{IR}$.
\item A high proportion (44\%) of X-ray detected sBzKs are heavily obscured AGN (HR $>-0.08$), while the remaining sBzKs are relatively unobscured AGN, with a few perhaps purely star-forming. The IR Excess sBzK and IR Non-Excess sBzK subsamples exhibit very similar proportions of heavily obscured AGN. Four of the X-ray detected sBzKs have been identified as Compton thick AGN from their X-ray spectra in the work of \citet{feruglio11} and \citet{brightman12b}. 
\item Broad band stacking of 688 X-ray undetected sBzKs, as well as IR Excess sBzK and IR Non-Excess sBzK subsamples, in the 4Ms CDFS and 2Ms CDFN yields soft stacked signals with effective $\Gamma\sim2$. Stacking of sBzKs binned by redshift and $L_{IR}$ produces similar results.
\item Narrow band stacking of sBzKs reveals a ``composite" spectrum with soft emission at low energies and hard emission at high energies. It is not possible to disentangle the emission from heavily obscured AGN, unobscured AGN and star-formation within the stacked signal. The resolution of the narrow band stacking is not sufficient to confirm or deny the presence of Iron K$\alpha$ line emission, so it cannot be established definitively whether or not a significant population of Compton thick AGN reside in the sBzKs. The IR Excess sBzKs and IR Non-Excess sBzKs have almost identical spectral shapes and, as with broad band stacking, there is no evidence that Compton thick AGN are more prevalent within IR Excess sBzKs than IR Non-Excess sBzKs.
\item The stacked emission of IR Excess sBzKs in the 1Ms CDFS is dominated by 6 IR Excess sBzKs lying just below the detection threshold at this depth, and these were subsequently detected in the deeper 4Ms CDFS. These sources are on average much harder (effective $\Gamma=0.7^{+0.2}_{-0.1}$) than the X-ray undetected sBzKs and are responsible for the hard stacked signal observed by \citet{daddi07b}.
\item Our results show that on the whole the discrepancy between 24$\mu$m  and UV-derived star formation in the BzK population is not due to AGN activity enhancing the 24$\mu$m flux. At high $L_{IR}$ (hence SFR) the 24$\mu$m overestimates the SFR, and the UV underestimates SFR for these BzK galaxies. The apparent lack of correspondence between X-ray emission from an AGN and 24$\mu$m excess is a puzzle which needs to be investigated in detail on an object-by-object basis. 
\item Passively evolving BzKs have an even higher X-ray detection rate ($\sim15\%$) than sBzKs ($\sim7\%$) and hence a high fraction of AGN. This is contradictory to the idea that black hole growth at $z\sim2$ is coeval with star-formation. The connection between the two appears to be purely circumstantial, rather than causal.  
\end{enumerate}

\section{Acknowledgements}
\label{sec: acknowledgements}
The authors are grateful to Prof Steve Warren and Dr Dave Clements for discussions and inputs that contributed significantly to this paper. CR acknowledges the financial support of STFC. We would also like to thank the referee for their helpful comments.

\pagebreak

\pagebreak

\begin{landscape}
\begin{table}
\begin{center}
\caption{Broad band stacking of X-ray undetected sBzK and pBzK galaxies in 4Ms CDFS and 2Ms CDFN. All fluxes calculated using flux conversion assuming $\Gamma=1.4$. Column(1): Sample of galaxies stacked; column (2): No. of galaxies used stacked; column (3): soft band S/N; column (4): hard band S/N; column (5): full band S/N; column (6): 0.5-2keV flux, derived from soft band stacked counts, units $10^{-17}$ erg cm$^{-2}$ s$^{-1}$, if we assume $\Gamma=2.0$ the soft band fluxes would be systematically 10\% lower; column (7): 2-10keV flux, derived from hard band stacked counts, units $10^{-17}$ erg cm$^{-2}$ s$^{-1}$; column (8): 0.5-10keV flux, derived from full band stacked counts, units $10^{-17}$ erg cm$^{-2}$ s$^{-1}$; column (9): HR, HR$=(H-S)/(H+S)$, where $H$ and $S$ are hard- and soft-band count rates respectively; column (10): effective $\Gamma$ of stack, derived from HR.}
\begin{tabular}{@{} l l c c c c c c c c}
\hline
Sample & No. Sources & S/N$_{0.5-2keV}$ & S/N$_{2-7keV}$ & S/N$_{0.5-7keV}$ & $F_{0.5-2keV}$ & $F_{2-10keV}$ & $F_{0.5-10keV}$ & HR & $\Gamma$\\
(1) & (2) & (3) & (4) & (5) & (6) & (7) & (8) & (9) & (10)\\
\hline
4Ms CDFS sBzKs & 370 & 16.00 & 4.24 & 12.97 & $0.51\pm0.03$ & $0.90\pm0.21$ & $1.71\pm0.13$ & $-0.53\pm0.08$ & $2.1^{+0.3}_{-0.2}$ \\
4Ms CDFS IRXs & 72 & 10.64 & 3.42 & 9.33 & $0.82\pm0.08$ & $1.66\pm0.49$ & $2.90\pm0.31$ &  $-0.48\pm0.11$ & $2.0^{+0.3}_{-0.3}$ \\
4Ms CDFS IRNXs & 298 & 12.40 & 3.02 & 9.75 & $0.43\pm0.03$ & $0.71\pm0.24$ & $1.43\pm0.15$ &  $-0.55\pm0.11$ & $2.2^{+0.4}_{-0.3}$ \\
2Ms CDFN sBzKs & 318 & 12.88 & 4.17 & 11.50 & $0.59\pm0.05$ & $1.22\pm0.29$ & $2.10\pm0.18$ & $-0.47\pm0.09$ & $2.0^{+0.2}_{-0.2}$ \\
2Ms CDFN IRXs & 107 & 10.34 & 3.31 & 9.34 & $0.88\pm0.09$ & $1.69\pm0.51$ & $3.08\pm0.33$ &  $-0.50\pm0.12$ & $2.0^{+0.4}_{-0.2}$ \\
2Ms CDFN IRNXs & 211 & 8.20 & 2.74 & 7.34 & $0.44\pm0.05$ & $0.98\pm0.36$ & $1.61\pm0.22$ &  $-0.45\pm0.15$ & $1.9^{+0.5}_{-0.3}$ \\
Combined sBzKs & 688 & 20.44 & 5.77 & 17.07 & $0.55\pm0.02$ & $1.04\pm0.18$ & $1.89\pm0.11$ & $-0.50\pm0.06$ & $2.0^{+0.2}_{-0.1}$ \\
Combined IRXs & 179 & 14.81 & 4.73 & 13.12 & $0.86\pm0.06$ & $1.68\pm0.36$ & $3.00\pm0.23$ & $-0.49\pm0.08$ & $2.0^{+0.3}_{-0.2}$ \\
Combined IRNXs & 509 & 14.84 & 3.95 & 12.08 & $0.43\pm0.03$ & $0.82\pm0.20$ & $1.50\pm0.13$ & $-0.51\pm0.09$ & $2.1^{+0.4}_{-0.2}$ \\
Combined pBzKs & 13 & 2.99 & 2.09 & 3.47 & $0.49\pm0.16$ & ... & $2.23\pm0.62$ &  ... & ... \\
\hline
\end{tabular}
\label{tab:xsstack}
\end{center}
\end{table}
\end{landscape}

\begin{landscape}
\begin{table}
\begin{center}
\caption{Broad band stacking of redshift and IR luminosity binned X-ray undetected sBzKs. Galaxies in 4Ms CDFS and 2Ms CDFN have been combined to boost the S/N ratio of the stacks. All fluxes calculated using flux conversion assuming $\Gamma\sim1.4$. Column(1): Sample of galaxies stacked; column (2): No. of galaxies used stacked; column (3): soft band S/N; column (4): hard band S/N; column (5): full band S/N; column (6): 0.5-2keV flux, derived from soft band stacked counts, units $10^{-17}$ erg cm$^{-2}$ s$^{-1}$; column (7): 2-10keV flux, derived from hard band stacked counts, units $10^{-17}$ erg cm$^{-2}$ s$^{-1}$; column (8): 0.5-10keV flux, derived from full band stacked counts, units $10^{-17}$ erg cm$^{-2}$ s$^{-1}$; column (9): HR, HR$=(H-S)/(H+S)$, where $H$ and $S$ are hard- and soft-band count rates respectively; column (10): effective $\Gamma$ of stack, derived from HR.}
\begin{tabular}{@{} l l c c c c c c c c}
\hline
Sample & No. Sources & S/N$_{0.5-2keV}$ & S/N$_{2-7keV}$ & S/N$_{0.5-7keV}$ & $F_{0.5-2keV}$ & $F_{2-10keV}$ & $F_{0.5-10keV}$ & HR & $\Gamma$\\
(1) & (2) & (3) & (4) & (5) & (6) & (7) & (8) & (9) & (10)\\
\hline
$1.2\le z <1.8$ & 340 & 12.62 & 4.06 & 10.91 & $0.38\pm0.03$ & $0.71\pm0.18$ & $1.29\pm0.12$ &  $-0.47\pm0.10$ & $2.0^{+0.3}_{-0.3}$\\
$1.8\le z <2.4$ & 304 & 14.28 & 4.12 & 11.97 & $0.47\pm0.03$ & $0.84\pm0.19$ & $1.54\pm0.13$ &  $-0.48\pm0.09$ & $2.0^{+0.3}_{-0.2}$\\
$2.4\le z <3.0$ & 97 & 6.46 & 0.62 & 4.31 & $0.36\pm0.05$ & ... & $1.01\pm0.21$ & ... & ... \\
$L_{IR} < 10^{11}$ & 108 & 5.96 & 2.12 & 5.31 & $0.33\pm0.06$ & $0.71\pm0.33$ & $1.16\pm0.21$ &  $-0.42\pm0.20$ & $1.8^{+0.7}_{-0.4}$\\
$10^{11} \le L_{IR} < 10^{12}$ & 393 & 11.16 & 3.60 & 9.46 & $0.29\pm0.03$ & $0.64\pm0.16$ & $1.01\pm0.11$ &  $-0.40\pm0.11$ & $1.8^{+0.3}_{-0.2}$\\
$10^{12} \le L_{IR} < 10^{13}$ & 203 & 13.42 & 3.73 & 11.31 & $0.56\pm0.04$ & $0.86\pm0.23$ & $1.78\pm0.16$ &  $-0.54\pm0.10$ & $2.1^{+0.4}_{-0.2}$\\
$L_{IR} \ge 10^{13}$ & 59 & 13.44 & 7.03 & 14.14 & $1.21\pm0.09$ & $2.89\pm0.47$ & $4.40\pm0.33$ &  $-0.37\pm0.08$ & $1.7^{+0.2}_{-0.1}$\\
\hline
\end{tabular}
\end{center}
\label{tab:z_ir_bin}
\end{table}
\end{landscape}

\begin{landscape}
\begin{table}
\begin{center}
\caption{Broad band stacking of X-ray undetected sBzK galaxies in 1Ms CDFS and sBzKs that were subsequently detected in the 4Ms CDFS. Galaxies in 4Ms CDFS and 2Ms CDFN have been combined to boost the S/N ratio of the stacks. All fluxes calculated using flux conversion assuming $\Gamma=1.4$. The 4Ms detected galaxies $F_{0.5-10keV}$ is lower than $F_{2-10keV}$ because the powerlaw ($\Gamma$=1.4) used to approximate the spectra is too soft. Column(1): Sample of galaxies stacked; column (2): No. of galaxies used stacked; column (3): soft band S/N; column (4): hard band S/N; column (5): full band S/N; column (6): 0.5-2keV flux, derived from soft band stacked counts, units $10^{-17}$ erg cm$^{-2}$ s$^{-1}$; column (7): 2-10keV flux, derived from hard band stacked counts, units $10^{-17}$ erg cm$^{-2}$ s$^{-1}$; column (8): 0.5-10keV flux, derived from full band stacked counts, units $10^{-17}$ erg cm$^{-2}$ s$^{-1}$; column (9): HR, HR$=(H-S)/(H+S)$, where $H$ and $S$ are hard- and soft-band count rates respectively; column (10): effective $\Gamma$ of stack, derived from HR.}
\begin{tabular}{@{} l l c c c c c c c c}
\hline
Sample & No. Sources & S/N$_{0.5-2keV}$ & S/N$_{2-7keV}$ & S/N$_{0.5-7keV}$ & $F_{0.5-2keV}$ & $F_{2-10keV}$ & $F_{0.5-10keV}$ & HR & $\Gamma$\\
(1) & (2) & (3) & (4) & (5) & (6) & (7) & (8) & (9) & (10)\\
\hline
1Ms CDFS sBzKs & 391 & 9.32 & 3.64 & 8.84 & $0.51\pm0.06$ & $1.29\pm0.36$ & $1.94\pm0.22$ & $-0.39\pm0.12$ & $1.8^{+0.3}_{-0.3}$ \\
1Ms CDFS IRXs & 80 & 5.56 & 3.60 & 6.32 & $0.70\pm0.13$ & $2.96\pm0.82$ & $3.20\pm0.51$ &  $-0.15\pm0.16$ & $1.3^{+0.4}_{-0.3}$ \\
1Ms CDFS IRNXs & 311 & 7.58 & 2.17 & 6.62 & $0.46\pm0.06$ & $0.86\pm0.40$ & $1.62\pm0.24$ &  $-0.51\pm0.17$ & $2.1^{+0.6}_{-0.4}$ \\
4Ms CDFS detected sBzKs & 12 & 9.69 & 9.96 & 13.74 & $2.41\pm0.25$ & $16.02\pm1.60$ & $13.89\pm1.01$ & $0.07\pm0.07$ & $0.8^{+0.2}_{-0.1}$ \\
4Ms CDFS detected IRXs & 6 & 7.52 & 8.57 & 11.29 & $2.72\pm0.36$ & $20.61\pm2.40$ & $16.98\pm1.50$ & $0.14\pm0.09$ & $0.7^{+0.2}_{-0.1}$ \\
4Ms CDFS detected IRNXs & 6 & 6.13 & 5.27 & 7.37 & $2.11\pm0.35$ & $11.15\pm2.12$ & $10.80\pm1.36$ &  $-0.04\pm0.12$ & $1.0^{+0.3}_{-0.2}$ \\
\hline
\end{tabular}
\end{center}
\label{tab:detstack}
\end{table}
\end{landscape}

\end{document}